\documentclass[a4paper,fleqn,usenatbib]{mnras}

\usepackage[T1]{fontenc}
\usepackage{ae,aecompl}

\usepackage{hyperref}
\usepackage{amsmath}	
\usepackage{cleveref}
\usepackage{graphicx}	
\usepackage{amssymb}	
\usepackage{listings}
\usepackage{multirow}
\usepackage[percent]{overpic}

\title[A study of four elusive binary pulsars in 47 Tuc]{Long-term observations of the pulsars in 47 Tucanae. I.\\A study of four elusive binary systems}

\author[A. Ridolfi et al.]{\parbox{\textwidth}{
A. Ridolfi,$^{1}$\thanks{E-mail: ridolfi@mpifr-bonn.mpg.de} 
P. Freire,$^{1}$
P. Torne,$^{1}$
C. O. Heinke,$^{1,2}$
M. van den Berg,$^{3,4}$
C. Jordan,$^{5}$
M. Kramer,$^{1,5}$
C. G. Bassa,$^{6}$
J. Sarkissian,$^{7}$
N. D'Amico,$^{8,9}$
D. Lorimer,$^{10}$
F. Camilo,$^{11}$
R. N. Manchester$^{7}$ and 
A. Lyne$^{5}$
}
\vspace{0.4cm} \\
\parbox{\textwidth}{
$^{1}$Max-Planck-Institut f\"ur Radioastronomie, Auf dem H\"ugel 69, D-53121 Bonn, Germany\\
$^{2}$Department of Physics, University of Alberta, CCIS 4-183, Edmonton, AB T6G 2E1, Canada\\
$^{3}$Anton Pannekoek Institute for Astronomy, University of Amsterdam, Science Park 904, 1098 XH Amsterdam, the Netherlands\\
$^{4}$Harvard$-$Smithsonian Center for Astrophysics, 60 Garden Street, Cambridge, MA 02138, USA\\
$^{5}$Jodrell Bank Centre for Astrophysics, School of Physics and Astronomy, The University of Manchester, Manchester M13 9PL, UK\\
$^{6}$ASTRON, the Netherlands Institute for Radio Astronomy, Postbus 2, 7990 AA, Dwingeloo, the Netherlands\\
$^{7}$CSIRO Astronomy and Space Science, Australia Telescope National Facility, Box 76, Epping, NSW 1710, Australia\\
$^{8}$Osservatorio Astronomico di Cagliari, INAF, via della Scienza 5, I-09047 Selargius (CA), Italy\\
$^{9}$Dipartimento di Fisica, Universit\`a degli Studi di Cagliari, SP Monserrato-Sestu km 0,7, 90042 Monserrato (CA), Italy\\
$^{10}$Department of Physics and Astronomy, West Virginia University, P.O. Box 6315, Morgantown, WV 26506, USA\\
$^{11}$Square Kilometre Array South Africa, Pinelands, 7405, South Africa
}}
\date{Accepted XXX. Received YYY; in original form ZZZ}

\pubyear{2016}

\begin{document}
\label{firstpage}
\pagerange{\pageref{firstpage}--\pageref{lastpage}}
\maketitle

\begin{abstract}
For the past couple of decades, the Parkes radio telescope has been regularly observing the millisecond pulsars in 47 Tucanae (47 Tuc). This long-term timing program was designed to address a wide range of scientific issues related to these pulsars and the globular cluster where they are located. In this paper, the first of a series, we address one of these objectives: the characterization of four previously known binary pulsars for which no precise orbital parameters were known, namely 47 Tuc P, V, W and X (pulsars 47 Tuc R and Y are discussed elsewhere).
We determined the previously unknown orbital parameters of 47 Tuc V and X and greatly improved those of 47 Tuc P and W. For pulsars W and X we obtained, for the first time, full coherent timing solutions across the whole data span, which allowed a much more detailed characterization of these systems. 47 Tuc W, a well-known tight eclipsing binary pulsar, exhibits a large orbital period variability, as expected for a system of its class. 47 Tuc X turns out to be in a wide, extremely circular, 10.9-day long binary orbit and its position is $\sim$\,3.8~arcmin away from the cluster center, more than three times the distance of any other pulsar in 47 Tuc. These characteristics make 47 Tuc X a very different object with respect to the other pulsars of the cluster.
\end{abstract}





\section{Introduction}
\label{sec:intro}

\newcommand{\dmunit}{pc\,cm$^{-3}$}
\newcommand{\msun}{M$_{\odot}$}
\newcommand{\rsun}{R$_{\odot}$}
\newcommand{\us}{$\mu$s}
\newcommand{\TO}{$T_0$}
\newcommand{\ergs}{~erg\,s$^{-1}$}

\newcommand{\chisquare}{\chi_{\textrm{red}}}

Globular clusters (GCs) are known to be host of a wealth of radio pulsars: with 146 currently known\footnote{\url{http://www.naic.edu/~pfreire/GCpsr.html}}, the pulsars in GCs account for more than 5\% of the total known pulsar population\footnote{\url{http://www.atnf.csiro.au/research/pulsar/psrcat} (version 1.54, \citealt{mht+05})} and a large fraction of millisecond pulsars (MSPs, here defined as those pulsars with spin periods of 10 ms or less).
\textcolor{black}{The population of pulsars in GCs also sharply contrasts with that of our Galaxy. The main reason is related to the much higher stellar densities that can be found in GCs, especially in their cores, which exceed those in the Galactic disk by several orders of magnitude. Such crammed environments provide the ideal conditions for the dynamical interaction of stars in two-body or even three-body encounters, the latter being the case of a star gravitationally interacting with a binary system. In this continuous exchange of gravitational energy, the more massive objects tend to sink towards the center of the cluster, a phenomenon called mass segregation \citep[see e.g. ][]{hge+05}. The process thus greatly fosters a concentration of neutron stars (NS) near the GC core and, consequently, the formation of exotic binary systems in which the NS can be spun up (or ``recycled'') by accreting matter and angular momentum from the companion star \citep{acr+82,rs82}. The observational evidence for this process is striking, as over 80\% of the GC pulsars known are MSPs and most of them reside close to the center of their host cluster. Nevertheless, there are substantial differences in the populations of pulsars in different GCs, particularly in the ratio between the isolated and the binary pulsars. As \citet{vf14} discussed, this ratio seems to be related to the encounter rate per binary of the GC: the higher the rate, the larger the fraction of isolated pulsars we are likely to see in the cluster.}

\textcolor{black}{47 Tucanae (also known as NGC 104 and, hereafter, 47~Tuc) is a well known GC visible in the southern sky.
The previous observations of 47 Tuc  with the Parkes radio telescope have provided outstanding scientific results that include the discovery of 25 radio MSPs \citep{mlj+90,mlr+91,rlm+95,clf+00,phl+16}, all of which have spin periods smaller than 8 ms. This is a very exceptional population, since only $\sim$\,7\% of the non-cluster pulsars in the Galaxy have spin periods of less than 8 ms.
As expected from a GC with a relatively low encounter rate per binary \citep{vf14}, a large fraction (60\%) of the pulsars in 47 Tuc are found in binary systems. This is similar to that for the Galactic non-cluster MSP population ($\sim$\,67\%), but very different from the general pulsar population, where the number in binaries is $\sim\,$7\%.
The binary pulsars in 47 Tuc include five ``Black Widow" pulsars (BWPs), with orbital periods of $P_b \sim 0.06 - 0.4$~days and companion masses of $M_{\rm c}~\sim0.03-0.05$~\msun, and at least two eclipsing ``Redback'' pulsars (RBs) with $P_b \sim 0.1 - 0.2$~days and $M_{\rm c} \sim0.1-0.5$~\msun (see e.g. \citealt{f05,r13} for reviews on BWPs and RBs)}.
The two categories share several common features, such as very short orbital periods (of the order of a few hours), highly recycled pulsars, very low mass companions that are undergoing mass loss, and, very often, regular eclipses in the pulsar radio signal. However, while in BWPs the companion stars have extremely low masses ($M_{\rm c} \lesssim 0.1$~\msun) and are being heavily ablated by the strong pulsar wind (which can make them almost completely stripped stars), in RBs the companions are mostly heavier ($M_{\rm c} \sim 0.1-0.5$~\msun), non-degenerate stars and   the mass loss is usually driven by a Roche-lobe overflow.

Of all the 47 Tuc pulsars, 17 have published timing solutions \citep{rlm+95,clf+00,fcl+01,fck+03,phl+16}, which allowed a detailed study of the dynamics of the cluster \citep{fcl+01}, the first detection of ionized gas in a globular cluster \citep{fkl+01}, the detection of all the pulsars at X-ray wavelengths \citep{gch+02,bgv05,bgh+06} and the optical detection of 5 of the 8 known MSP$-$White-Dwarf (WD) systems \citep[\textcolor{black}{namely} 47 Tuc Q, S, T, U and Y;][]{egh+01,cpf+15,rvh+15}.

\textcolor{black}{Since the previous publication of radio timing data by \citet{fck+03}}, we have continued timing the millisecond pulsars in 47 Tuc. The main objectives of this long-term timing programme are diverse, and include: a) avoiding the loss of phase-connection \textcolor{black}{in the timing} of any of the pulsars; b) monitoring orbital phase variations for the BWP and RB binaries; \textcolor{black}{this is particularly important not only to understand their unpredictable orbital dynamics, but also in relation to the previous point, since losing track of the precise orbital phase can easily cause a loss of phase-connection in the timing;} c) improving the orbital measurements for the stable MSP$-$WD binaries, in particular the measurement of relativistic and kinematic effects; d) improving the determination of the proper motions of all pulsars, in order to study the internal dynamics of the cluster; e) determining orbits and timing solutions for previously known binary pulsars that still lack them; f) discovering new pulsars.
This is part of a series of papers that will present the results of this \textcolor{black}{long-term programme}. In the work of \citet{phl+16} and in the upcoming papers, we address the search for new pulsars.
In \textcolor{black}{the present} paper, we have used 16 years of archival data taken at the Parkes radio telescope to further characterize four among the faintest binary systems of 47 Tuc, namely pulsars P, V, W and X, for which an accurate orbital solution (and, hence, also a timing solution) was not known. Accurate orbital and timing solutions for pulsars R and Y will instead be presented in Freire et al. (in preparation).

In Section 2 we describe our dataset. In Sections 3 we explain the two search methods that we have used to re-detect the four pulsars and we give refined orbital parameters for all of them. In Section 4 we describe our timing analysis, whose results are presented \textcolor{black}{and discussed} in Section 5. In Section 6 we summarize our findings.

\vskip 0 cm
\section{Observations}
\label{sec:data_reduction}
The dataset used consisted of 519 pointings to 47 Tuc done with the 64-m Parkes radio telescope, at 414 different epochs\footnote{\textcolor{black}{We call ``pointing'' a single, continuous, observation of 47 Tuc, that hence corresponds to a single recorded data file. An ``epoch'' is instead used as a synonym for ``day''. The full list of pointings with the relative observing parameters can be found at: \url{http://www3.mpifr-bonn.mpg.de/staff/pfreire/47Tuc/Observations_Table.html}}}, between October 1997 and August 2013. All the observations together amounted to a total of $\sim$\,1770 hours. Of the 519 pointings, 438 were longer than 1 hour, whereas 30 were shorter than 10 minutes\textcolor{black}{, these latter mostly being the result of accidental failures in the observing system.}

The observing setup was mostly the same as described in \citet{fck+03}.
The vast majority of the observations were carried out at a frequency of $\simeq$\,1.4 GHz using the central beam of the Parkes Multi-Beam Receiver \citep[PMB,][]{swb+96}, whose size (full-width half-maximum, FWHM) is 14.4 arcmin.
Occasionally, when the PMB was not available, we observed 47 Tuc with the H-OH receiver (FWHM = 14.8~arcmin) at the \textcolor{black}{same frequency.}
Another handful of observations were made
with other receivers, namely the ``436-MHz'' \textcolor{black}{(FWHM~=~45~arcmin), the ``660-MHz'' (FWHM~$\simeq$~30~arcmin)}, the \textcolor{black}{10-50cm} (FWHM = 6.4~arcmin at 10 cm, FWHM~=~30~arcmin at 50 cm) and the METH6 receiver (FWHM = 3.4 arcmin).
Until August 1999 we used the PMB filterbank as back-end. When using the PMB receiver, we had a central observing frequency $f_c = 1374$ MHz, a bandwidth of 288 MHz divided into 96, 3-MHz wide, channels and a sampling time of 125 $\mu$s. At that frequency, for a pulsar with the average dispersion measure (DM) of 47 Tuc ($\simeq$\,24.4~\dmunit), the dispersive smearing is 234$\,\mu$s across a single channel, resulting in an effective time resolution of 265$\,\mu$s.
From August 1999 we switched to a different configuration, and most of the observations were carried out with the Analogue Filter Bank (AFB) back-end, with a central frequency $f_c = 1390$~MHz, 256 MHz of bandwidth divided into 512, 0.5-MHz wide, channels and a sampling time of 80 $\mu$s. For this observing system the dispersive smearing of the 47 Tuc pulsars is only 37 $\mu$s across \textcolor{black}{a single channel}, which gives an effective time resolution of 88 $\mu$s.
After summing the signals of the two polarizations and integrating for the time of the chosen sampling interval, all the data were 1-bit digitized and then recorded onto magnetic tapes as filterbank files.

\section{SEARCHES AND ORBITAL DETERMINATION}

\subsection{Accelerated search}
We have carried out a deep search of the whole dataset using a machine endowed with two NVIDIA K20 and two NVIDIA K40 graphics processing units (GPUs) and a search pipeline that made use
of the GPU-based version\footnote{\url{https://github.com/jintaoluo/presto2\_on\_gpu}} of
the \texttt{PRESTO}\footnote{\url{http://www.cv.nrao.edu/~sransom/presto}} pulsar search package \citep{r01}.
Each observation was split into chunks of both 20 and 60 minutes of length. On each chunk we first ran the \texttt{rfifind} routine to look for bursts of radio frequency interference (RFI), creating a mask and filtering out bad frequency channels or time intervals. Then, \textcolor{black}{again for each chunk}, a 0-DM (i.e. without correcting for the dispersion effect due to the interstellar medium) time series was created and searched for prominent persistent terrestrial low-level RFI (dubbed as ``birdies'') with a Fast Fourier Transform. The periodicities found were marked as interference, to be ignored during the analysis, and stored in a so-called ``zaplist'' file.   After that, nine de-dispersed time series were created for an assumed DM of 23.90 \dmunit{ to} 24.70 \dmunit{ with} steps\footnote{Such interval more than covered the DM range of the pulsars known in 47 Tuc when the search was done. However, a newly discovered pulsar, 47 Tuc ab, has a DM of $\simeq 24.94$ {\dmunit } \citep{phl+16}. \textcolor{black}{This implies that the range of DM's in 47 Tuc is larger than what we assumed in the present work. Our current search effort, to be reported elsewhere, takes this into account, by significantly broadening the range of DM searched.}} of 0.10 \dmunit. The time series were then searched in the Fourier domain by the \texttt{accelsearch} routine, summing up to 8 harmonics, with a $z$-value\footnote{The $z$-value is defined as $z=T_{\textrm{obs}}^2 a_l/(cP)$ where $T_\textrm{obs}$ is the observation length, $a_l$ is the pulsar acceleration along the line of sight, $c$ is the speed of light, and $P$ is the pulsar spin period. When looking for binaries, \texttt{accelsearch} searches $\pm z$ Fourier bins around a \textcolor{black}{given} frequency.} ranging from 0 (absence of acceleration) up to 1200 for the 60-min chunks and up to 200 for the 20-min chunks. All the candidate pulsar-like signals with a detection significance of $\sigma \geq 2.0$ were recorded onto files, one for each chunk and each DM.

Once the search was done, we looked for all the candidates with the highest $\sigma$ values that were detected at periods close to the known
barycentric spin periods of the pulsars of our sample of interest (namely 47 Tuc P, V, W and X) and that appeared at multiple DM values.
Table \ref{tab:pulsar_detections} shows the number of detections that we obtained for each pulsar.

\begin{table}
\begin{center}
\caption{\textcolor{black}{Number of detections of the pulsars obtained with the three different search methods as well as by \textcolor{black}{refolding} the whole dataset with the relative timing solution (reported in Table \ref{tab:timing_solutions}).}}
\label{tab:pulsar_detections}
\begin{tabular}{lcccc}
\hline
					      	&   \texttt{PRESTO}		& \texttt{PRESTO}			&  					& \textcolor{black}{Folding}\\
Pulsar						&    (20-min)			& (60-min) 					&  \TO-search		& \textcolor{black}{(Timing)}\\
\hline
47 Tuc P						& 	0						&0								&	4		 &	\textcolor{black}{4}	    \\			
47 Tuc V						& 	0						&0								&	7		&	\textcolor{black}{19}		\\		
47 Tuc W						&	1						&1								& 	23		&	\textcolor{black}{34}		\\			
47 Tuc X						&  12							&16								& 	$-$	&	\textcolor{black}{312}		\\
\hline
\end{tabular}
\end{center}
\end{table}

\begin{figure}
\centering
	\includegraphics[width=\columnwidth]{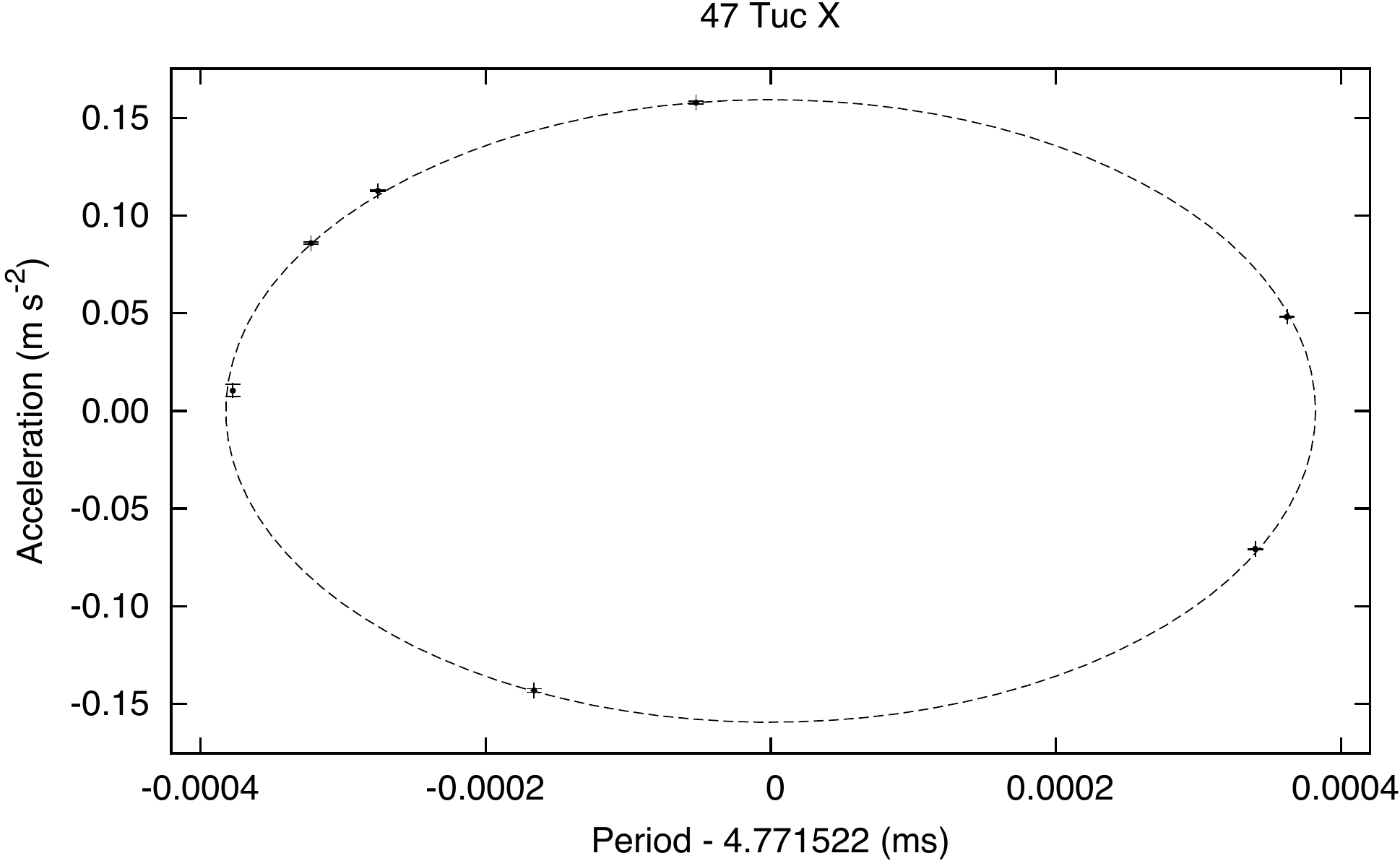}
  	\caption{Period-acceleration diagram for 47 Tuc X. The fit returned an orbital period of $P_b \simeq 10.9$~days and a projected semi-major axis of $x_p \simeq 11.9$ lt-s for the pulsar orbit.}
  	\label{fig:47TucX_Per_Acc}
\end{figure}

\subsection{Orbital solution for 47 Tuc X}

47 Tuc X is a formerly known, but unpublished, millisecond pulsar that had been discovered in previous searches of the cluster\footnote{The discovery of 47 Tuc X will be discussed in Freire et al. (in preparation).}. The pulsar had already been found with non-zero accelerations, indicating the occurrence of a binary motion. 
Our  \texttt{PRESTO} 20- and 60-min search pipelines produced a sufficiently large number of new detections of this pulsar to allow us to estimate its orbital parameters with the \mbox{period-acceleration} diagram method, as described in \citet{fkl01}.
Among the 16 unique epochs at which the pulsar was detected, we selected the 7 best ones. We then used the \texttt{TEMPO}\footnote{\url{http://tempo.sourceforge.net}} pulsar timing software to fit these latter for the barycentric spin period, $P$, and its first derivative, $\dot{P}$, thus finding a local timing solution for each epoch. After converting the $\dot{P}$ values into radial accelerations $a_l$ (through the relation $a_l = c\dot{P}/P$, where $c$ is the speed of light), we fitted for the \textcolor{black}{orbital period, $P_b$, and projected semi-major axis of the pulsar orbit, $x_p$,} via eq. (7) and (8) of \citet{fkl01}, using the \texttt{circorbit}\footnote{\textcolor{black}{\texttt{circorbit} {is part of a suite of codes developed at Jodrell Bank, whose list can be found at \url{http://www.jb.man.ac.uk/pulsar/Resources/tools.html}. The \texttt{circorbit} code can be downloaded from \url{http://www3.mpifr-bonn.mpg.de/staff/pfreire/programs/circorbit.tar.}}}} code.
The data points and best-fit curve are shown in Fig. \ref{fig:47TucX_Per_Acc}. The fit revealed that the pulsar is in a circular orbit \textcolor{black}{with $P_b \simeq 10.9$~days and $x_p \simeq 11.9$ lt-s}.

\begin{figure*}
\centering
  \includegraphics[width=0.325\textwidth]{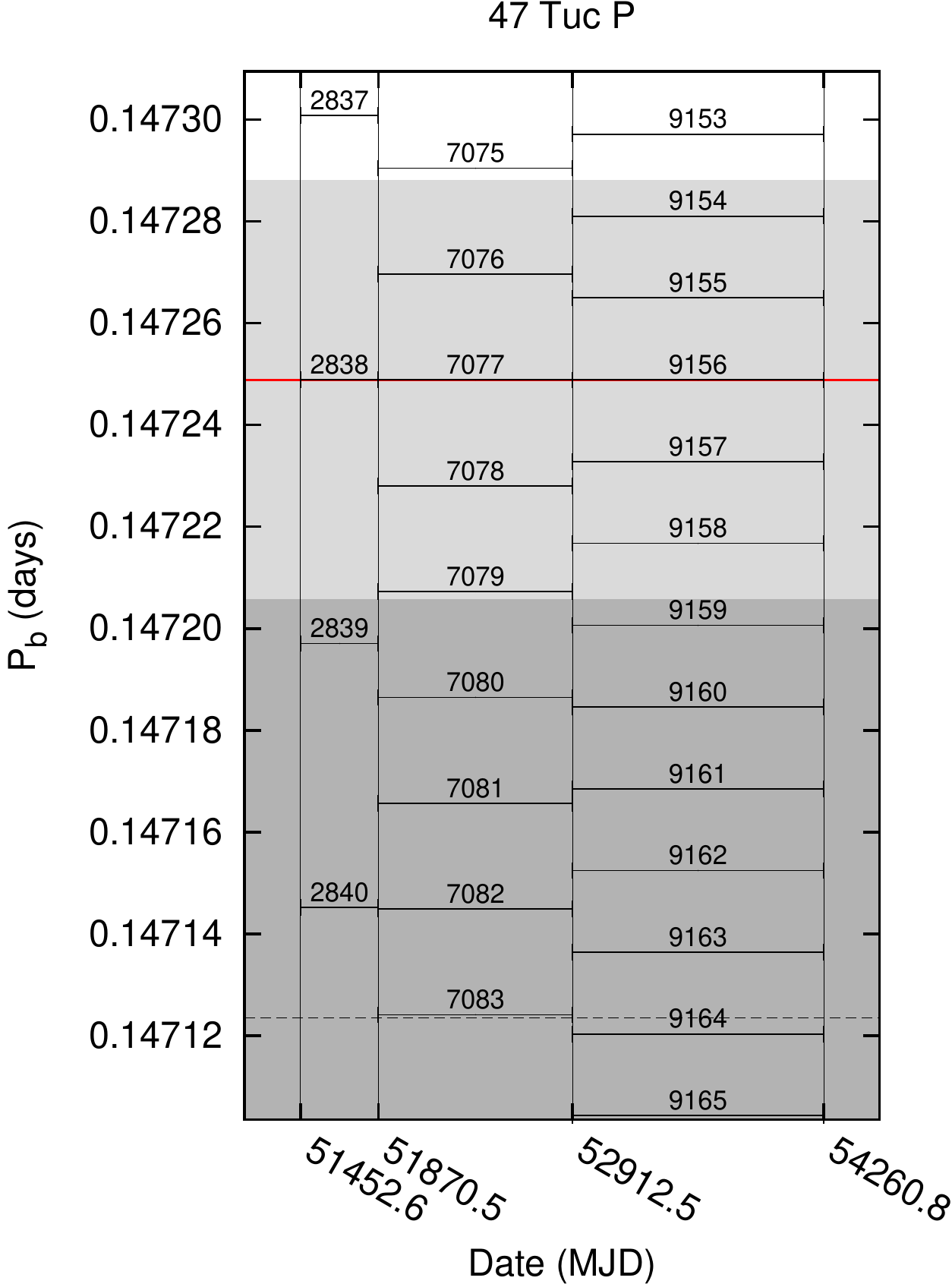}
  \includegraphics[width=0.32\textwidth]{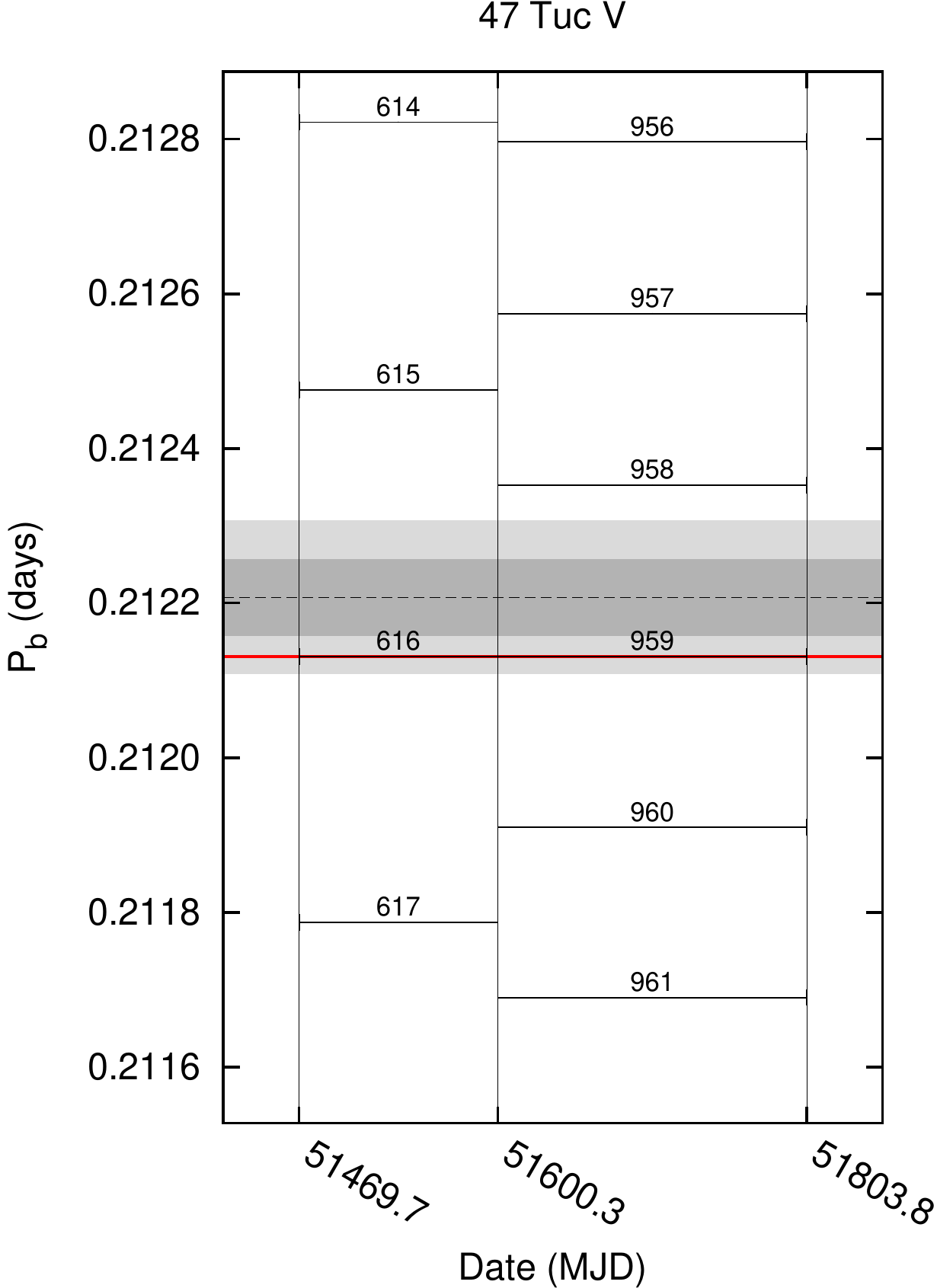}
  \includegraphics[width=0.32\textwidth]{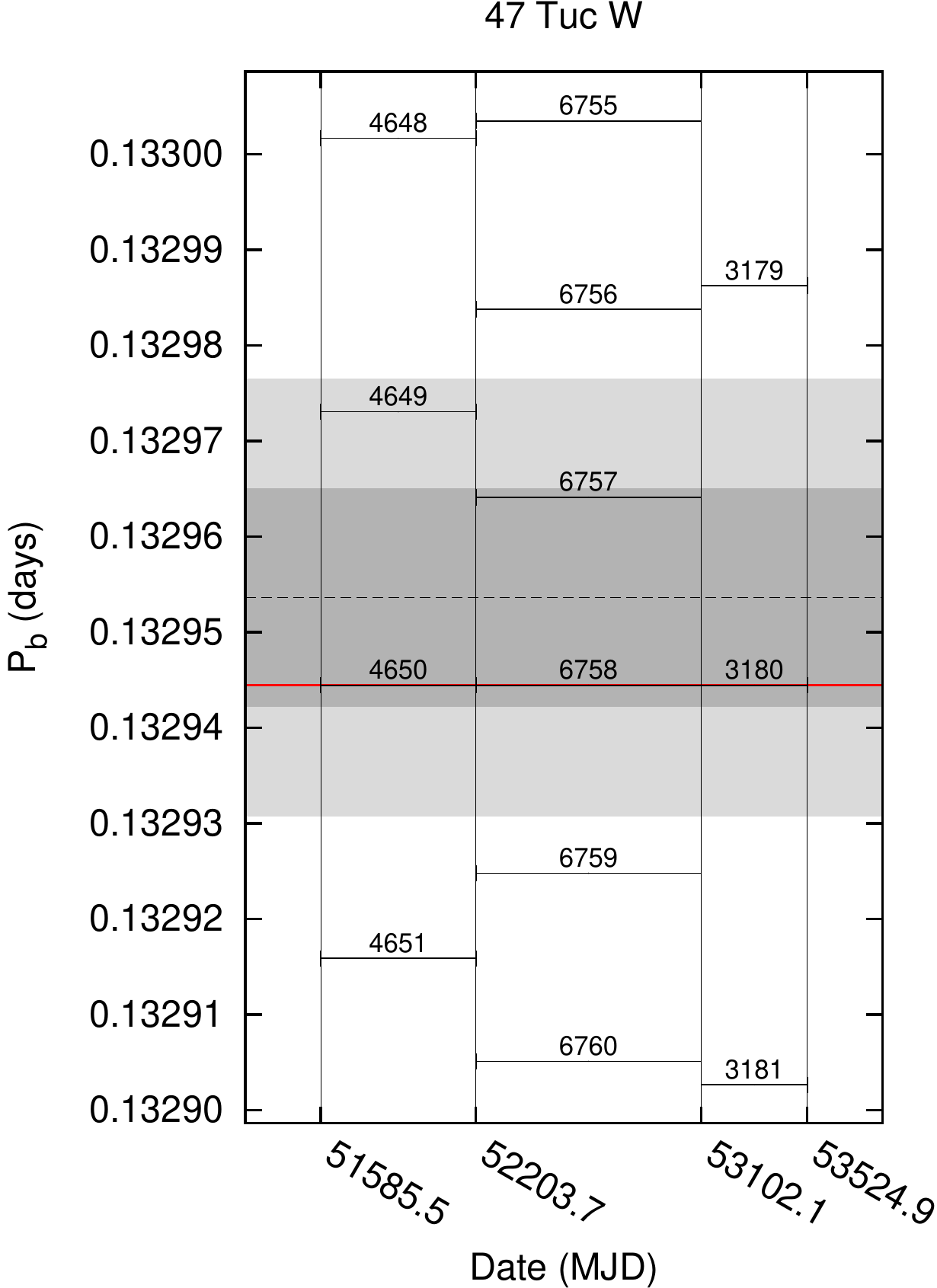}
  \caption{\textcolor{black}{Periodograms for pulsars P, V and W. In each plot the dashed line marks the initial guess of $P_b$, corresponding to the best-fit value derived by running \texttt{TEMPO} on the best detection obtained from our \TO-search; the dark grey and light grey areas represent the relative formal $1\sigma$ and $2\sigma$ uncertainties, respectively.}  The solid vertical lines indicate the measured \TO{ values} that we used. The horizontal segments indicate the orbital period obtained by dividing the time interval that they connect by the integer number reported above them. In each plot, the red line highlights the orbital period that is an integer submultiple of all time intervals.}
  \label{fig:periodograms}
\end{figure*}

\subsection{$T_0$-search}
As shown in Table \ref{tab:pulsar_detections}, pulsars P, V and W were either detected once or not at all in our \texttt{PRESTO} searches. The reasons for the very low rate of detection are manifold: these pulsars are all intrinsically very faint, their orbital periods are very short (reducing the maximum possible integration time where the constant acceleration approximation is effective, see e.g. \citealt{rce03}) and two of them (V and W) are eclipsed for a large fraction of the orbit. Only when both the scintillation and the observed orbital phase interval were favorable, did we have the chance of detecting these pulsars.
Because of their very compact orbits, comparable with the length of a single observation, the orbital parameters of pulsars P, V, and W were already known at the time of their discovery with a reasonable degree of accuracy. However, the uncertainties on their orbital periods, together with the large orbital variability that some of them show, prevented us from predicting their correct orbital phases at a later time.  Unlike the uncertainty associated with the projected semi-major axis, the uncertainty $\Delta \phi_b$(t) on the orbital phase $\phi_b$ at any time $t$, grows linearly with time, as:

\begin{equation}
\Delta \phi_b (t) = \frac{\Delta P_b \cdot (t - t_0)}{P_b^2}
\end{equation} 
where $t_0$ is the epoch at which the orbital period $P_b$ and its uncertainty $\Delta P_b$ were estimated. The uncertainty on $\phi_b$ translates into an uncertainty in \TO, the epoch of passage at periastron\footnote{For circular orbits (as is the case for all the pulsars discussed in this paper), where the periastron is not defined, \TO\ is conventionally chosen to coincide with $T_{\rm asc}$, the epoch of passage at the ascending node.}, in the pulsar ephemerides. This is the reason why, even knowing the spin periods and the rough values of the orbital parameters of the three pulsars at the epoch of their discoveries, we could not simply fold the  data to look for possible new detections: the correction for the R\o mer delay due to the binary motion would be applied with a wrong orbital phase, thus smearing out the signal.

What we did instead was a search in orbital phase, or, equivalently, in the value of \TO. For this reason we will refer to it as \TO-search throughout the rest of the paper.

First, we created a full-length (i.e. this time we did not split the observations into chunks) de-dispersed time series for each observation of our 47 Tuc dataset, at a DM of 24.40 \dmunit. We chose to use a single value of the DM since, for the purpose of just detecting the pulsars, a more precise value is unimportant.  
Then, for each pulsar, we folded all the de-dispersed time series with an ephemeris that had all the known parameters fixed, with the exception of \TO. The latter was instead searched, allowing its value to vary between the epoch of the beginning of the observation, $t_\textrm{start}$, and $t_\textrm{start} + P_b$, where $P_b$ was the (rough) value of the orbital period of the pulsar considered.
The choice of using de-dispersed time series instead of the original filterbank files is justified by the huge advantage in computational speed that the former have over the latter, when folding. To further minimize the computational costs without degrading the sensitivity of our search, the choice of the step size was also crucial and will be now discussed in detail.

\subsubsection{Choice of the step size}
Let us consider the case of a perfectly circular orbit (as is practically  the case for pulsars P, V and W) and let us assume to know the exact orbital period $P_b$ and projected size of the orbit $x_p$. Then, if the  orbital phase predicted by an incorrect ephemeris differs from the actual phase $\phi_b$ by a quantity $\Delta \phi_b$, the applied correction for the orbital R{\o}mer delay at a time $t$ will be wrong by the quantity:
\begin{equation}
\begin{split}
x_p \sin[\phi_b(t) + \Delta \phi_b] - x_p \sin[{\phi_b}(t)] =\\
A \sin\Bigg[ \frac{2 \pi}{P_b}(t - T_0) + \frac{\Delta \phi_b}{2}\Bigg]
\end{split}
\end{equation}
where the amplitude of the sinusoid is:
\begin{equation}
A = 2 x_p\sin(\Delta \phi_b/2)
\end{equation}

When folding, the integrated pulse profile will have a width  $W\sim A$. If $A\gtrsim P$, where $P$ is the pulsar spin period, the residual orbital modulation will completely smear out the pulsar signal, resulting in a non-detection.

For this reason, we imposed that the step size in our \TO-search be sufficiently small such that, in the case of the best trial value, $A \leq P/4$, which translates into the condition:
\begin{equation}
\Delta \phi_b   \leq 2\arcsin\Bigg( \frac{P}{8 x_p} \Bigg).
\end{equation}
The resulting total number of steps is thus:
\begin{equation}
N = \frac{2 \pi}{ \Delta \phi_b} = \frac{\pi} {\arcsin(P/8 x_p)}.
\end{equation}
For pulsars P, V and W, this latter amounts to 262, 4441 and 2601, respectively.

As already mentioned, this calculation is valid under the assumption of knowing the correct values for $P_b$ and $x_p$. In general, this is not the case, since our goal is in fact to \emph{obtain} a better value for $P_b$. Incorrect starting values of $P_b$ and $x_p$ introduce additional errors to the applied R{\o}mer delay corrections, which further degrade our sensitivity. This degradation can be partly compensated by using a finer grid of values in the \TO-search.
In our work, in the case of pulsars P and W, we started with very good values of $P_b$ and $x_p$, precise enough to fold an observation correctly, once the exact value of \TO\ was determined; for V, instead, the values were not as precise.
In our analysis, in order to maximize our sensitivity, we decided to use a number of steps which was $\sim$\,2 times larger than the values reported above for all the three pulsars.

\subsection{Periodograms and improved orbital periods for 47 Tuc P, V and W}

After folding all the observations with all the trial values of \TO, we ordered by signal-to-noise ratio (S/N) the resulting archives, that were then inspected visually.
From every archive where we clearly detected the pulsar \textcolor{black}{(see Table \ref{tab:pulsar_detections} for the number of detections obtained with the \TO-search for each pulsar)}, we extracted the topocentric pulse Times-of-Arrival (ToAs) and fitted for \TO{ only} with \texttt{TEMPO}, to obtain a refined value of it for each epoch. These latter were in turn used to improve the measurement of the orbital period  using the periodogram method described by \citet{fkl01}, which essentially consists in finding a value for $P_b$ that fits an integer number of times between any pair of the measured values of \TO. Fig. \ref{fig:periodograms} shows the periodograms for 47 Tuc P, V and W.

\section{Radio timing}
\label{sec:radio_timing}
\begin{figure}
\centering
  \includegraphics[width=\columnwidth]{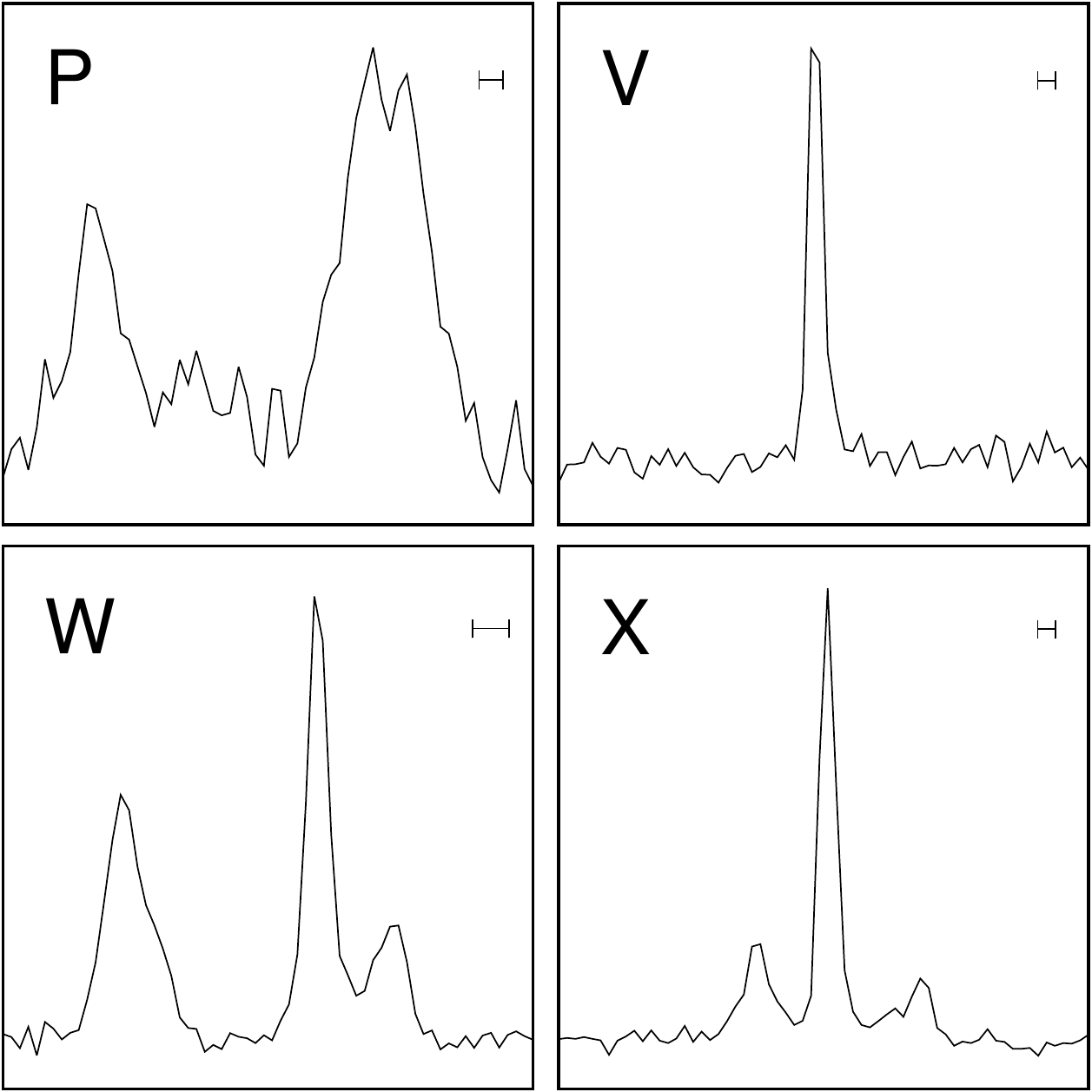}
  \caption{Integrated pulse profiles of the four pulsars studied. Each profile is shown over one full rotation. The horizontal bar on the top right of each panel shows the nominal AFB back-end sampling interval (80 $\mu$s).}
  \label{fig:pulse_profiles}
\end{figure}

For each pulsar, we constructed a first ephemeris that contained the average barycentric (i.e. referred to the Solar System Barycenter, SSB) spin period, the nominal position and proper motion of the cluster, as well as a simple Keplerian (BT) binary model \citep{bt76} with the refined orbital parameters obtained through the period-acceleration diagram for pulsar X, and through the \TO-search and the periodogram method for P, V and W.
We then used the ephemeris to fold the entire dataset with \texttt{DSPSR} \citep{vb11}. This occasionally allowed us to obtain a few additional detections. After that, we used the \texttt{PSRCHIVE} pulsar software \citep{hvm04,vdo12} to extract topocentric ToAs, by cross-correlating the pulse profiles against a high-S/N template that we previously built from the best detections (see \citealt{t92} for the description of the method). Integrated pulse profiles for all the four pulsars are shown in Fig. \ref{fig:pulse_profiles}. After applying the clock corrections of the Parkes radio telescope, the ToAs were converted by \texttt{TEMPO} and referred to the SSB. This was achieved by first correcting for the Earth rotation using data from the International Earth Rotation Service, and then correcting for the Earth motion around the SSB using the JPL DE421 Solar System \textcolor{black}{Ephemeris} \citep{fwb09}. The differences between the observed ToAs and the  values predicted by our preliminary \textcolor{black}{pulsar} ephemeris were then fitted by \texttt{TEMPO} through a weighted $\chi^2$ minimization for the astrometric, rotational and orbital parameters, as well as for initial arbitrary jumps between groups of ToAs belonging to different epochs.
The first resulting non-coherent timing solution was then used to refold the data again, often yielding new detections of the relative pulsar and, thus, more ToAs that were then included in the fit. The whole procedure was iterated a few times until we converged to the best solution. For two of the pulsars (namely 47 Tuc W and X) we obtained a phase-connected timing solution, which is thus capable of unambiguously account for every single rotation of the source.  The best-fit timing parameters for all the pulsars are reported in Table \ref{tab:timing_solutions}.

\begin{figure*}
\centering
  \includegraphics[width=0.49\textwidth]{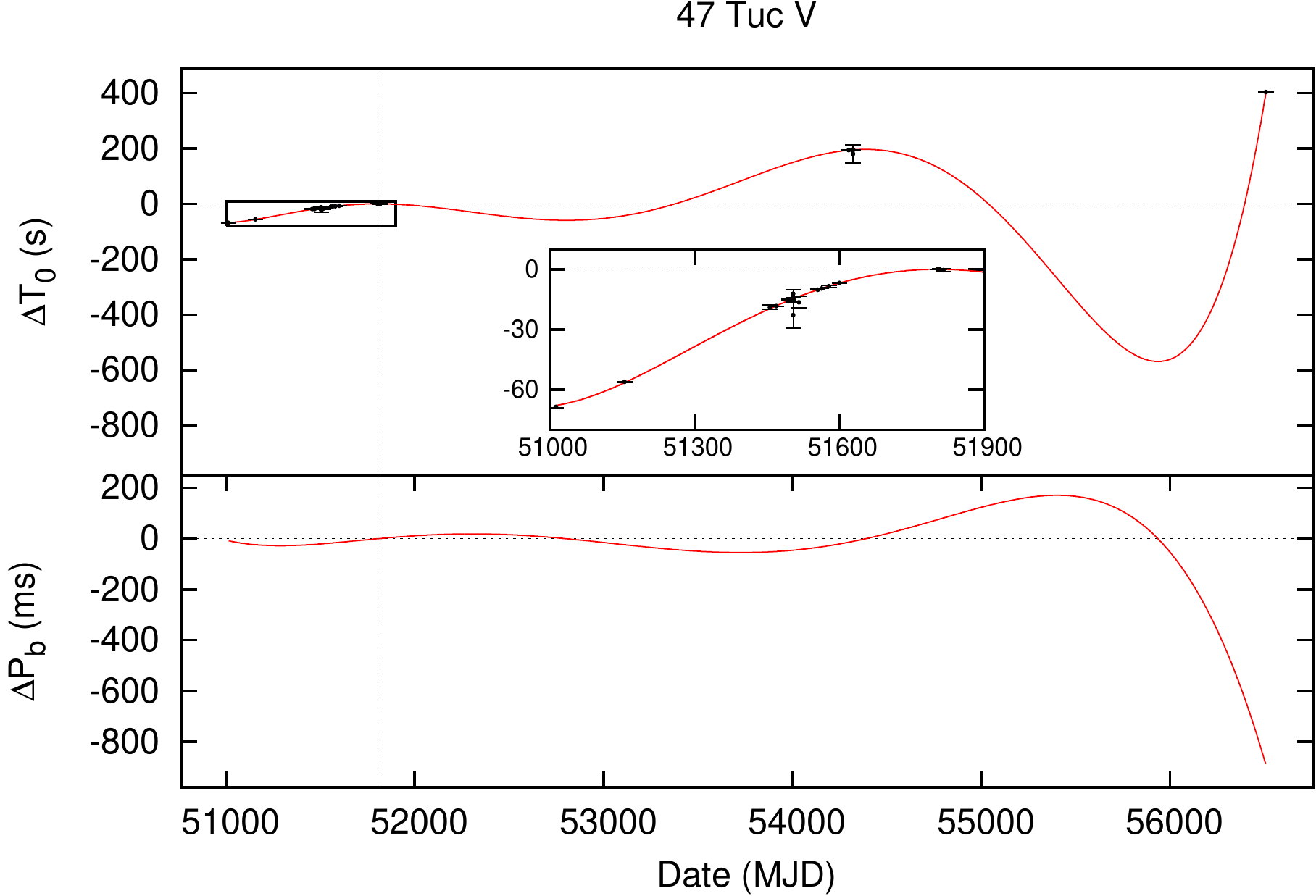}
  \includegraphics[width=0.48\textwidth]{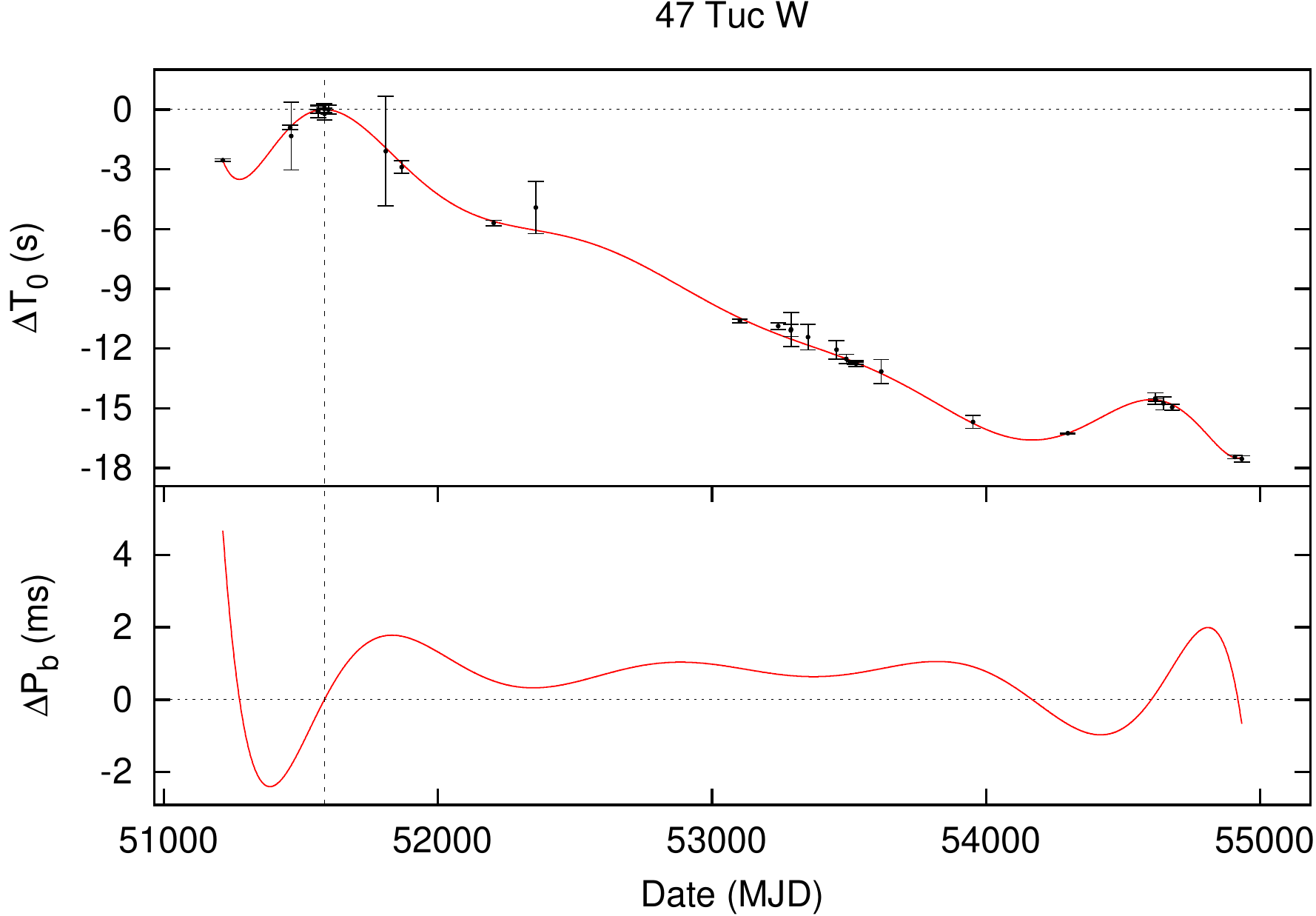}
    \caption{Orbital variability of pulsar V (left) and W (right). Top panels: deviation of the epoch of passage at the ascending node from the prediction of a standard Keplerian model as a function of time. The red line is the theoretical prediction of the BTX model, the black points are the values measured with timing (see text for the detailed description of the method). Bottom panels: corresponding change of the orbital period from the reference value as a function of time. Note that this latter is just the time derivative of the red line of the above panel. }
  \label{fig:orbital_variability}
\end{figure*}

\section{Results and Discussion}
\subsection{47 Tuc P}
\label{sec:47TucP}
47 Tuc P is a 3.64-ms binary pulsar in a 3.5-h orbit that was discovered in a single observation (on MJD 50689) by \citet{
clf+00}. The authors determined the binary parameters by extracting 14 ToAs that covered more than one orbit.
Our \TO-search yielded an additional 4 detections (of which only one had high S/N and the other three were relatively faint) from which we were able to extract another 14 ToAs, thus bringing the total number to 28, over a time span of $\sim$\,9.8 years. 
Because of the sparsity of the data, we were unable to obtain a phase-connected timing solution. Nevertheless, we were able to fit for the spin period, spin period derivative, position and  binary parameters. We kept the proper motion fixed to the value of global motion of the cluster. The latter was calculated as an unweighted mean of the proper motions of all the 22 pulsars with a timing solution available\footnote{All the updated timing solutions of the pulsars in 47 Tuc and a more thorough analysis of their proper motions will be presented in Freire et al. (in preparation). }, and resulted in $\mu_\alpha = 4.9 \pm 0.9$~mas~yr$^{-1}$ in right ascension and  $\mu_\delta =  -2.7 \pm 0.7$~mas~yr$^{-1}$ in declination. We preferred an unweighted mean over a weighted mean to avoid a global proper motion biased by the motion of the pulsars with the most precise timing measurements, as well as to obtain more realistic uncertainties.

\begin{table*}
\caption{Timing parameters for pulsars P, V, W and X as obtained from fitting the observed ToAs with \texttt{TEMPO}. All the uncertainties reported are $1\sigma$ and are the nominal values calculated by \texttt{TEMPO}.}
\label{tab:timing_solutions}
\begin{center}{\scriptsize
\setlength{\tabcolsep}{8pt}
\renewcommand{\arraystretch}{1.2}
\begin{tabular}{l c c c c}
\hline
Pulsar  &    47 Tuc P                                                             &   47 Tuc V                                                             &    47 Tuc W                                                             &    47 Tuc X\\
\hline
\multicolumn{5}{c}{Observation and Data Reduction Parameters}  \\
\hline\hline
Phase-connected solution?                                             \dotfill &   No                                                      		&   No                                                            		&   Yes                                                       		&   Yes  \\
Reference epoch (MJD)                                                 \dotfill &   51000.000                                                           &   51000.000                                                           	&   50000.000                                                           	&   54000.000  \\
Start of timing data (MJD)                                            \dotfill &   50689.609                                                           &   51012.862                                                              	&   51214.216                                                              	&   50981.881  \\
End of timing data (MJD)                                              \dotfill &   54260.850                                                           &   56508.781                                                              	&   54934.047                                                              	&   56508.949\\
Solar System Ephemeris                                                \dotfill &   DE421                                                               &   DE421                                                                  	&   DE421                                                                  	&   DE421\\
Terrestrial time standard                                             \dotfill &   UTC(NIST)                                                           &   UTC(NIST)                                                              	&   UTC(NIST)                                                              	&   UTC(NIST)\\
Time Units                                                            \dotfill &   TDB                                                                 &   TDB                                                                    	&   TDB                                                                    	&   TDB \\
Number of ToAs                                                        \dotfill &   28                                                                  &   99                                                                     	&   199                                                                    	&   719\\
Residuals r.m.s. ($\mu$s)                                              \dotfill &   10.43                                                               &   25.32                                                                  	&   10.20                                                                  	&   14.51\\
Binary Model                                                          \dotfill &   BT                                                                  &   BTX                                                                    	&   BTX                                                                    	&   ELL1\\
\hline
\multicolumn{5}{c}{Fitted Timing Parameters}  \\
\hline\hline
Right ascension, $\alpha$ (J2000)                                     \dotfill &   00:24:20(29)                                                        &   00:24:05.359$^{\textrm{a}}$                               		&   00:24:06.058(1)                                                       	&   00:24:22.38565(9)\\
Declination, $\delta$ (J2000)                                         \dotfill &   -72:04:10(62)                                                       &   -72:04:53.20$^{\textrm{a}}$                                             &   -72:04:49.088(2)                                                       	&   -72:01:17.4414(7)\\
Proper Motion in $\alpha$, $\mu_\alpha$ (mas yr$^{-1}$)    	   \dotfill &   4.9$^{\textrm{b}}$						&   4.9$^{\textrm{b}}$                                                      &   6.1(5)                                                 		&   5.8(1)\\
Proper Motion in $\delta$, $\mu_\delta$ (mas yr$^{-1}$)               \dotfill &   -2.7$^{\textrm{b}}$                                              	&   -2.7$^{\textrm{b}}$                                                     &   -2.6(3)                                                               	&   -3.3(2)\\
Parallax (mas)                                                        \dotfill &   0.2132$^{\textrm{c}}$						&   0.2132$^{\textrm{c}}$                                                   &   0.2132$^{\textrm{c}}$                                                 	&   0.2132$^{\textrm{c}}$\\
Spin Frequency, $f$ (s$^{-1}$)                                        \dotfill &   274.49748(2)                                                        &   207.892963(3)                                                          	&   425.10779625320(5)                                                    	&   209.576694635350(2)\\
1st Spin Frequency derivative, $\dot{f}$ (Hz s$^{-2}$)                \dotfill &   -5(3)$\times 10^{-14}$                                              &   -                                                                      	&   1.56415(2)$\times 10^{-14}$                                           	&   -8.0646(3)$\times 10^{-16}$\\
Dispersion Measure, DM (pc cm$^{-3}$)                                 \dotfill &   24.29(3)                                                            &   24.105(8)                                                              	&   24.367(3)                                                             	&   24.539(5)\\
Projected Semi-major Axis, $x_p$ (lt-s)                               \dotfill &   0.038008(4)                                              		&   0.74191(2)                                                             	&   0.243443(2)                                                           	&   11.9170570(9)\\
Orbital Eccentricity, $e$                                             \dotfill &   0                                                           	&   0                                                           		&   0                                                           		&   --\\
Longitude of Periastron, $\omega$ (deg)                               \dotfill &   0                                                         		&   0                                                        		&   0                                                         		&   --\\
Epoch of passage at Periastron, $T_0$ (MJD)                           \dotfill &   52912.5481(1)                                                       &   51803.775137(2)                                                        	&   51585.3327393(2)                                                       	&   --\\
\textcolor{black}{First Laplace-Lagrange parameter, $\eta$}                                                          \dotfill &   --                                                                  &   --                                                                     	&   --                                                                     	&   4(1)$\times 10^{-7}$\\
\textcolor{black}{Second Laplace-Lagrange parameter, $\kappa$}                                                          \dotfill &   --                                                                  &   --                                                                     	&   --                                                                     	&   -2(2)$\times 10^{-7}$\\
Epoch of passage at Asc. Node, $T_\textrm{asc}$ (MJD)            \dotfill &   --                                                                     &   --                                                                     &   --                                                                     	&   53278.0247041(2)\\
Orbital Period, $P_b$ (days)                                          \dotfill &   0.147248891(7)                                                         &   $-$                                                                     &   $-$                                                                     &   10.921183545(1)\\
Orbital Period derivative, $\dot{P}_b$ (10$^{-12}$ s s$^{-1}$)        \dotfill &   $-$                                                                     &   $-$                                                                     &   $-$                                                                     &   6(2)\\
Orbital Frequency, $f_b$ (s$^{-1}$)                                   \dotfill &   $-$                                                                     &   5.45609205(4)$\times 10^{-5}$                                          &   8.70594798(3)$\times 10^{-5}$                                          &   $-$                                                                     \\
1st Orbital Freq. derivative, $f^{(1)}_b$ (s$^{-2}$)                  \dotfill &   $-$                                                                     &   -2.34(4)$\times 10^{-18}$                                              &   -1.26(4)$\times 10^{-18}$                                              &   $-$\\
2nd Orbital Freq. derivative, $f^{(2)}_b$ (s$^{-3}$)                  \dotfill &   $-$                                                                     &   2.34(6)$\times 10^{-26}$                                               &   4.0(2)$\times 10^{-26}$                                                &   $-$\\
3rd Orbital Freq. derivative, $f^{(3)}_b$ (s$^{-4}$)                  \dotfill &   $-$                                                                     &   2.31(6)$\times 10^{-33}$                                               &   6.3(3)$\times 10^{-33}$                                                &   $-$\\
4th Orbital Freq. derivative, $f^{(4)}_b$ (s$^{-5}$)                  \dotfill &   $-$                                                                     &   -6.7(2)$\times 10^{-41}$                                               &   -9.2(4)$\times 10^{-40}$                                               &   $-$\\
5th Orbital Freq. derivative, $f^{(5)}_b$ (s$^{-6}$)                  \dotfill &   $-$                                                                     &   5.6(1)$\times 10^{-49}$                                                &   6.3(3)$\times 10^{-47}$                                                &   $-$\\
6th Orbital Freq. derivative, $f^{(6)}_b$ (s$^{-7}$)                  \dotfill &   $-$                                                                     &   $-$                                                                     &   -2.7(1)$\times 10^{-54}$                                               &   $-$\\
7th Orbital Freq. derivative, $f^{(7)}_b$ (s$^{-8}$)                  \dotfill &   $-$                                                                     &   $-$                                                                     &   7.4(3)$\times 10^{-62}$                                                &   $-$\\
8th Orbital Freq. derivative, $f^{(8)}_b$ (s$^{-9}$)                  \dotfill &   $-$                                                                     &   $-$                                                                     &   -1.22(5)$\times 10^{-69}$                                              &   $-$\\
9th Orbital Freq. derivative, $f^{(9)}_b$ (s$^{-10}$)                 \dotfill &   $-$                                                                     &   $-$                                                                     &   9.3(4)$\times 10^{-78}$                                                &   $-$\\
\hline
\multicolumn{5}{c}{Derived Parameters}  \\
\hline\hline
Spin Period, $P$ (ms)                                          		\dotfill &   3.6430207(2)                                                         &   4.81016762(7)                                                                     &   2.3523445319370(3)                                                                     &   4.77152291069355(5)                                                        \\
1st Spin Period Derivative, $\dot{P}$ (s s$^{-1}$)                                          \dotfill &      7(4)$\times 10^{-19}$                                                      &   $-$                                                                     &   -8.6553(1)$\times 10^{-20}$                                                                     &    1.83609(7)$\times 10^{-20}$                                                        \\
Mass Function, $f(M_p, M_c)$ (\msun)                                          \dotfill &   2.7188$\times 10^{-6}$                                                         &   9.743$\times 10^{-3}$                                                                    &   8.764$\times 10^{-4}$                                                                     &   1.524$\times 10^{-2}$                                                        \\
Minimum Companion Mass$^{d}$, $M_c^{\rm min}$ (\msun)                                        \dotfill &   0.0176                                                         &   0.3048                                                                     &   0.1269      											 &   0.3616   \\
Distance from Cluster Center, $\theta_\perp$ (arcmin)                                        \dotfill &   $-$                                                          &   $-$                                                                     &   0.087      											 &   3.828   \\
Distance from Cluster Center$^{e}$, $\theta_c$ (core radii)                                        \dotfill &   $-$                                                         &   $-$                                                                     &   0.242      											 &   10.633   \\
\hline
\multicolumn{5}{p{16.5cm}}{$^{a}$The position was fixed to the nominal center of 47 Tuc as reported in the SIMBAD database (\url{http://simbad.u-strasbg.fr/simbad/sim-id?Ident=47+Tuc}). $^{b}$The proper motion was set to the average value of the 22 pulsars with a phase-connected timing solution available. $^{c}$The value of the parallax was fixed to the value corresponding to a cluster distance of 4.69 kpc \citep{wgk+12}. $^{d}$Assuming a pulsar mass of $M_p= 1.4$\,\msun. $^{e}$Assuming a core radius of 0.36 arcmin, as reported in the Harris GC catalogue (\url{http://www.physics.mcmaster.ca/~harris/mwgc.dat}).  } 
\end{tabular} }
\end{center} 
\label{tab:timing_solutions}
\end{table*}

Both the position and spin period derivative are, as expected, rather poorly determined and their values should be taken as just indicative. 
The orbit is correctly described by a simple Keplerian model, with no evidence of orbital period derivatives. This, together with the very low mass of the companion ($M_{\rm c} \sim 0.02$~\msun{ for} a pulsar mass of 1.4 \msun{ and} an orbital inclination $i = 60^\circ$) and the total absence of radio eclipses, strengthens the hypothesis that 47 Tuc P is a Black-Widow type system, possibly seen at a low orbital inclination. Indeed, as pointed out by \citet{f05}, some BWPs have lower mass functions as a result of lower orbital inclinations, which in turn make the display of eclipses less likely. On the contrary, systems seen more edge-on appear to have larger masses and tend to exhibit eclipses.

To measure the DM, we summed in time each of our 4 detections, retaining as many frequency sub-bands as possible, according to the S/N. From every sub-band we extracted one time of arrival, for a total of 12 usable ToAs. The latter were then fitted for the DM only, again allowing arbitrary phase jumps between the different epochs. The resulting measured DM was $24.29 \pm 0.03$ \dmunit, a value slightly lower than the average, which suggests that the pulsar is on the near side of the cluster. In fact, we can obtain a rough estimate of the distance component along the line of sight, $d_\parallel$, through the linear relation found by \citet{fkl+01} between the DM and the radial distance from the plane of the sky containing the center of 47 Tuc. Using this relation and the measured DM, we infer $d_\parallel \simeq -1.36$ pc, with the negative sign indicating that the pulsar is indeed placed between the observer and the cluster center.

\textcolor{black}{No calibration data are currently available to determine the flux densities of the 47 Tuc pulsars. However,} having the lowest detection rate ($ 0.97$\%) in our dataset, 47 Tuc P is  probably the faintest currently known pulsar in the cluster. 
\subsection{47 Tuc V}
47 Tuc V is a 4.81-ms binary pulsar in a 5.1-h orbit, first presented by \citet{clf+00}. The pulsar was detected twice in their data (on MJD 51012 and 51055), and the authors were able to give an estimate of the orbital parameters by extracting a total of 11 ToAs. At both epochs the pulsar was on the near side of its orbit, as the spin period was increasing during the observations.
With the \TO-search, we were able to detect 47 Tuc V \textcolor{black}{in another 7 pointings}. Through our timing analysis as described in Section \ref{sec:radio_timing} we were able to obtain an additional 12 (mostly faint) new detections, bringing the total number to 19. After constructing a high-S/N template profile, we carefully extracted 88 usable new ToAs. In total, we had 99 ToAs spanning $\sim$\,10.7 years.
As for pulsar P, we were unable to obtain phase-connection with our timing: in addition to the proper motion (set to the same average value as for P), the large covariances seen also forced us to not fit for the spin period derivative as well as for the position, the latter being set to the nominal cluster center value. However, allowing arbitrary jumps between groups of ToAs, we derived an incoherent timing solution that enabled us to measure the spin period and the binary parameters with good precision. The parameters are reported in Table~\ref{tab:timing_solutions}.

Using the same method as described in the previous section, we independently measured the DM, obtaining a value of $24.105 \pm 0.008$~\dmunit, the lowest amongst all the 47 Tuc pulsars.  
For this DM value, the projected distance along the line of sight from the cluster center, predicted by the linear relation of \citet{fkl+01}, is $d_\parallel \simeq -4.12$ pc.

\subsubsection{Long-term orbital variability}
\label{sec:orbital_variability_V}
When timing 47 Tuc V, a simple BT model turned out to be inadequate to correctly describe the orbit and it was instead necessary to use a BTX model (D. Nice, unpublished; \mbox{\url{http://tempo.sourceforge.net}}), which allows to fit for multiple orbital frequency derivatives. In short, the number of orbits $N_b^\textrm{\tiny{BTX}}$ at any given time $t$ is described by a Taylor expansion:

\begin{equation}
\label{eq:BTX_model}
N_b^\textrm{\tiny{BTX}}(t) = N_0 +\sum_{k=0}^{N_d} \frac{1}{(k+1)!}f^{(k)}_b (t -T_0)^{(k+1)}
\end{equation}
where $N_0$ is an arbitrary constant, $N_d$ is the number of orbital frequency derivatives, $T_0$ is the time of passage at periastron (or at the ascending node, in the case of a circular orbit), and $f^{(k)}_b$ is the value of the $k$-th orbital frequency derivative calculated at the time $t=T_0$.
In the case of 47 Tuc V, in order to achieve a reduced chi-square $\chisquare \sim 1$ in the timing residuals, we had to introduce five orbital frequency derivatives.

To study how well the BTX model is actually describing the data, we followed a method similar to the one used by \citet{nbb+14} for PSR J1731$-$1847. 
For each group of ToAs (that is, for each detection), we calculated the  epoch of passage at periastron, $T_0^\textrm{pred}$, closest to the mid-point $\langle T_\textrm{obs} \rangle$ of the observation, that a simple Keplerian model would predict. This is equivalent to finding $t$ closest to $\langle T_\textrm{obs} \rangle$ in Eq. (\ref{eq:BTX_model}), after setting to zero all the orbital frequency derivatives.
Then, we fitted the ToAs with \texttt{TEMPO} for $T_0$ only, using a simple Keplerian model (i.e., setting all the $f^{(k)}_b$ to zero) in the ephemeris, and keeping all the other parameters fixed, thus obtaining a measured value $T_0^\textrm{obs}$. The difference $\Delta T_0 = T_0^\textrm{pred} - T_0^\textrm{obs}$ was then plotted against the sum of the terms with $k\geq 1$ of Eq. (\ref{eq:BTX_model}), which represents the deviations from the Keplerian model. The result can be seen in the top-left panel \textcolor{black}{of} Fig. \ref{fig:orbital_variability}; in the lower-left panel, we show the consequent predicted change in the orbital period as a function of time.
\textcolor{black}{The strong orbital variability is a common characteristic among Redback binary systems and it thus supports the hypothesis that 47 Tuc V belongs to this class, as already hinted by the mass function and the presence of eclipses (see Section \ref{sec:eclipses_V})}.

 It is important to note that, although the 5-derivative BTX model used here is able to predict the orbital phase of 47 Tuc V observations within \textcolor{black}{the time spanned by the data}, it has no predictive capacity outside that range. Also, the predicted large swing in $\Delta T_0$ between MJD\,$\sim$\,54500 and $\sim$\,56700 has no data supporting it. For these reasons, we should not consider this BTX binary model as a faithful representation of the binary orbit evolution, especially in those time intervals where the data points are very sparse. Rather, the model should be seen as a display of the unpredictable orbital variability similar to that observed in other Redback systems.

\begin{figure*}
	{\begin{overpic}[width=\textwidth]{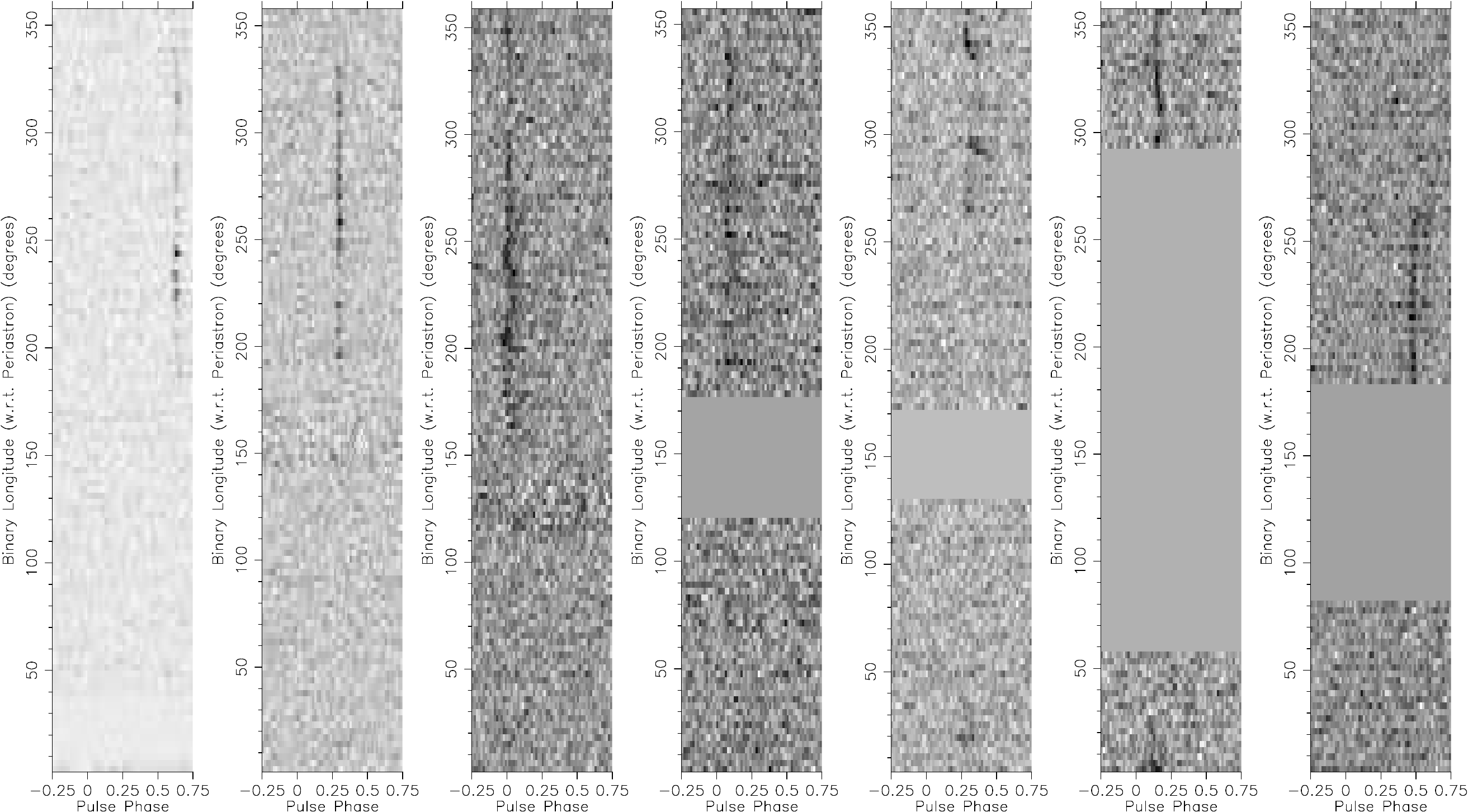}
\put (8.0,55.7) {\footnotesize{1}}
\put (22.3,55.7) {\footnotesize{2}}
\put (36.6,55.7) {\footnotesize{3}}
\put (50.8,55.7) {\footnotesize{4}}
\put (65.2,55.7) {\footnotesize{5}}
\put (79.5,55.7) {\footnotesize{6}}
\put (93.8,55.7) {\footnotesize{7}}
\end{overpic} }
   \caption{Intensity as a function of pulse phase ($x$-axis) and orbital phase $\phi_b$ ($y$-axis) for 7 sample detections of 47 Tuc V. As can be clearly seen, the pulsar is always eclipsed for about 50\% of its orbit around the superior conjunction (i.e. $\phi_b = 0.25$). In addition, several short-lived eclipses often occur at other orbital phases, sometimes accompanied by excess time delays (panel 5) and sometimes with a broadening of the pulse (panel 3). }
   \label{fig:eclipses_V}
 \end{figure*}

\subsubsection{Eclipses}
\label{sec:eclipses_V}
In the two discovery observations, \citet{clf+00} noted that the pulsar signal was being irregularly eclipsed over time scales as short as their 2-min long sub-integrations. Our new 19 detections allowed us to get a deeper insight into the morphology and, possibly, nature of the eclipses. Not only did we confirm the presence of irregular short-lived eclipses, we also found that the pulsar is \emph{always} invisible for roughly 50\% of the orbit around its superior conjunction, likely being enshrouded by the gas that the companion star is losing.

The presence of long and persistent eclipses implies that the system inclination, $i$, cannot be small and that the companion is likely non-degenerate. Choosing a conservative lower limit of $i \gtrsim 20^{\circ}$ and a pulsar mass of $M_p = 1.4 $~\msun, the mass function implies that the companion has a mass in the range $ 0.30\textrm{ \msun} \lesssim M_{\rm c} \lesssim 1.17 \textrm{ \msun}$, pointing towards a main sequence type. This range, together with the observed eclipsed fraction of the orbit of $\Delta \phi_b^\textrm{ecl}\sim 0.5$, can be used to estimate the rough projected size, $R_c$, of a supposedly spherical obscuring gas cloud through simple geometrical considerations. In the case of a circular, edge-on ($i=90^{\circ}$) orbit, this can be expressed by the relation:
\begin{equation}
R_c = a_p (1 + q) \sin \bigg( \frac{2 \pi \Delta \phi_\textrm{b}^\textrm{ecl} }{2} \bigg)
\end{equation}
where $a_p$ is the radius of the pulsar orbit and \mbox{$q = M_p/M_{\rm c}$} is the mass ratio.
From the previous considerations, \mbox{$1.19 \lesssim q \lesssim 4.59$}, and thus $1.79 \textrm{ \rsun}\lesssim R_c \lesssim 2.05 \textrm{ \rsun}$. On the other hand, the inferred Roche-lobe radius $R_L$ of the companion, calculated with the approximate formula by \citet{e83}, is in the range $ 0.75 \textrm{ \rsun} \gtrsim R_c \gtrsim  0.46 \textrm{ \rsun}$, implying that the gas is extending far beyond the gravitational influence of the companion star. Hence, the companion is probably undergoing continuous mass loss through the Roche-lobe overflow of its outer layers. 
We confirm the presence of frequent, short-lived, eclipses that hide the pulsar over time scales of minutes when the pulsar is \textcolor{black}{on} the near side of its orbit (panels 1-2 of Fig. \ref{fig:eclipses_V}). Sometimes the signal disappeared even for relatively long times (panels 5 and 7), whereas in another occasion (panel 5), the pulse also underwent a visible delay of roughly $\sim$\,0.5 ms.
Very likely, besides the presence of a large gas cloud surrounding the companion, clumps of ionized plasma of smaller size are wandering around the binary, occasionally intercepting the pulsar signal. When the physical characteristics (density and temperature) of the gas clump are such that the optical depth at our observing frequency is low, we still see the signal, but delayed because of the additional \textcolor{black}{contribution to the} dispersion measure that the clump introduces. Given our observing frequency of 1390 MHz, in the case of the 0.5-ms delay seen in panel 5 of Fig. \ref{fig:eclipses_V}, the inferred extra DM is $\simeq 0.24$ \dmunit.

 \begin{figure*}
\centering
  \includegraphics[width=0.9\textwidth]{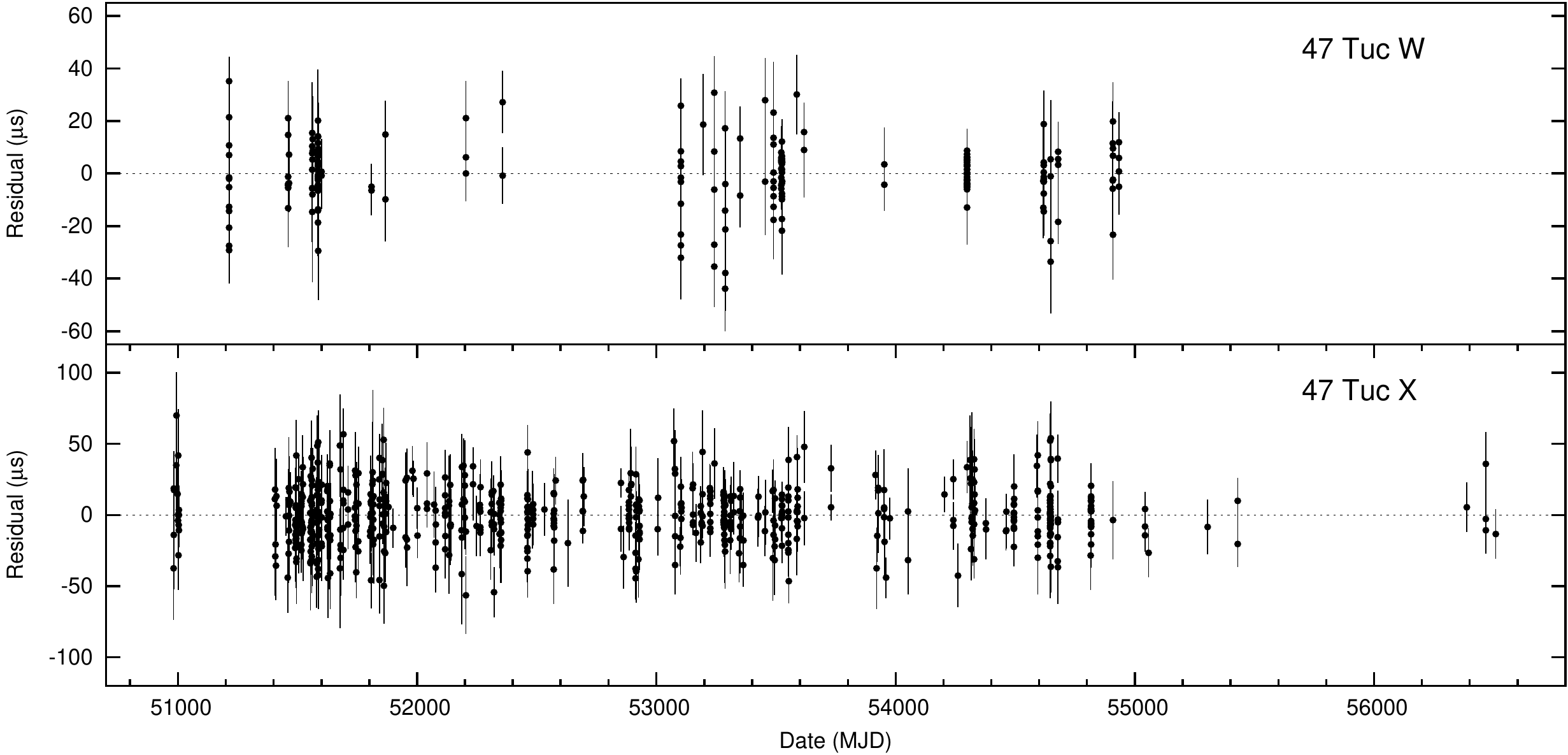}
  \caption{Timing residuals of 47 Tuc W (top panel) and 47 Tuc X (bottom panel).}
  \label{fig:timing_residuals}
\end{figure*}
\subsubsection{A transitional MSP?}
As evident from Fig. \ref{fig:orbital_variability}, the detections of 47 Tuc V appear clustered in three main groups, separated by two large gaps of $\sim$\,6.8 (MJD 51815$-$54298) and $\sim$\,6.0 years (MJD 54322$-$56508), respectively, where the pulsar was never visible.
A similar behavior has recently been seen in the so-called ``Transitional Millisecond Pulsars" (tMSPs), a new subclass of Redback binary MSPs that periodically switch between radio-MSP and Low-Mass X-ray Binary (LMXB) phases, over a time scale of a few years. During the latter phase, the neutron star accretes matter from the companion, radio pulsations disappear and the system is much brighter at X-ray, optical, and sometimes $\gamma$-ray wavelengths. 
The archetype of this class is PSR J1023+0038, an eclipsing binary millisecond pulsar which was initially identified as a cataclysmic variable by \citet{bwb+02}, due to its optical spectrum that indicated the presence of an accretion disk. Subsequently, \citet{ta05} proposed that the system was actually a LMXB, a hypothesis also supported by \citet{hsc+06}. A few years later, the system was seen changing to a radio-MSP state \citep{asr+09}, and then switching back to a LMXB state \citep{sah+14,dmm+15}.
The other two currently known tMSPs are PSR J1824$-$2452I in the globular cluster M28 \citep{pfb+13} and the newly discovered PSR~J1227$-$4853 \citep[XSS~J12270$-$4859 in its LMXB state, ][]{dfb+10, dbf+13, bph+14, rrb+15}.

The strong orbital variability of 47 Tuc V, its persistent and irregular eclipses, the disappearance of its radio pulsations within intervals of a few years, and the mass of the companion of the order of $\sim$\,0.3$-$0.4~\msun, are all typical characteristics of tMSPs that suggest that pulsar V might indeed be a new transitional millisecond pulsar.
To constrain this hypothesis, we have looked into archival \textcolor{black}{\emph{Chandra}} data and the literature on optical studies, to search for signs of an active tMSP.
We used two methods: with the \textcolor{black}{\emph{Chandra}} data, we searched for X-ray outbursts; from the optical studies, we searched for periodic signals at 47 Tuc V's orbital period. 

Transitional MSPs, during their active LMXB states, have X-ray luminosities in the range of $L_{\rm X} = 10^{33}-10^{34}$\ergs; such luminous objects would be easily detected by \textcolor{black}{\emph{Chandra}} anywhere in 47 Tuc.
The data considered are grouped into four widely separated epochs.
The first group of \textcolor{black}{\emph{Chandra}} observations took place on MJD 51619$-$51620 \citep{ghe+01}. These observations were only 19 days after a detection of pulsations from 47 Tuc V on MJD 51600. Thus, it seems unlikely that pulsar V would be X-ray active during this \textcolor{black}{\emph{Chandra}} observation, and the X-ray sources at 10$^{33}$\ergs\ in this observation can be ruled out.
The next two groups of observations were carried out on  MJD 52547$-$52558 and 53723$-$53743, both in the midst of the first gap in radio pulsation. No new \textcolor{black}{\emph{Chandra}} sources significantly brighter than $10^{31}$\ergs\ were seen in these data. 
   The final \textcolor{black}{\emph{Chandra}} observations occurred in the MJD range 56905$-$57055, after the radio data considered, and also show no new bright sources. We point out that there are at least two caveats that must be considered: a) it is possible that  pulsar V was X-ray active during small fractions of the time in the radio detection gaps; b) it is possible that the actual position of 47 Tuc V is beyond the field of view of \textcolor{black}{\emph{Chandra}}.
Even though both hypotheses appear unlikely, we are currently not able to rule them out.

We make reference to three major optical surveys for periodicities in 47 Tuc: two ground-based, and one using the {\em Hubble Space Telescope (HST)} over a smaller field of view.
The optical light curve of PSR~J1023+0038, which has a similar orbital period (4.75\,h), though slightly lower companion mass (0.24\,M$_\odot$) compared to 47~Tuc~V, varies between a magnitude of $V_\mathrm{min}=17.0$ and $V_\mathrm{max}=16.4$ in its active state 
\citep{hgs+13}, and between $V_\mathrm{min}=17.7$ and $V_\mathrm{max}=17.35$ in its passive state \citep{ta05}. Placing PSR~J1023+0038 ($d=1.37$~kpc,
\citealt{dab+12}) at the distance of 47 Tuc would suggest 
$V_\mathrm{min}=19.7$ and $V_\mathrm{max}=19.1$ in the active state, or $V_\mathrm{min}=20.3$ and $V_\mathrm{max}=20.0$ in the passive state.
The $B$- and $V$-band photometric variability study of 47 Tuc done by \citet{krp+13} covered the entire Parkes beam down to about $V=20.5$, though the core region could not be studied due to confusion. These observations were obtained primarily in 2009 and 2010 and coincide with the second gap in which 47~Tuc~V is not detected, suggesting it may have been in the active LMXB state of a tMSP. In the LMXB state, the tMSPs PSR~J1023+0038 and XSS~J12270$-$4859 have sinusoidal light curves \citep{cbc+14,mpb+15}, i.e. having a single minimum and maximum per orbital cycle. None of the variables in the study by \citet{krp+13} had a period comparable to the \textcolor{black}{5.1-h} orbital period of 47~Tuc~V.
The range where \citet{krp+13} found variables with similar orbital periods (the faintest variable detected had $V_{\rm max}=19.6$) covered the range expected for 47~Tuc~V in the active state. This suggests that, if 47~Tuc~V was an active tMSP in 2009-2010 with similar properties as PSR\,J1023+0038, it is likely that it would have been detected by the authors, \textcolor{black}{unless 47~Tuc~V resides in the core region.}

 \begin{figure}
   \includegraphics[width=\columnwidth]{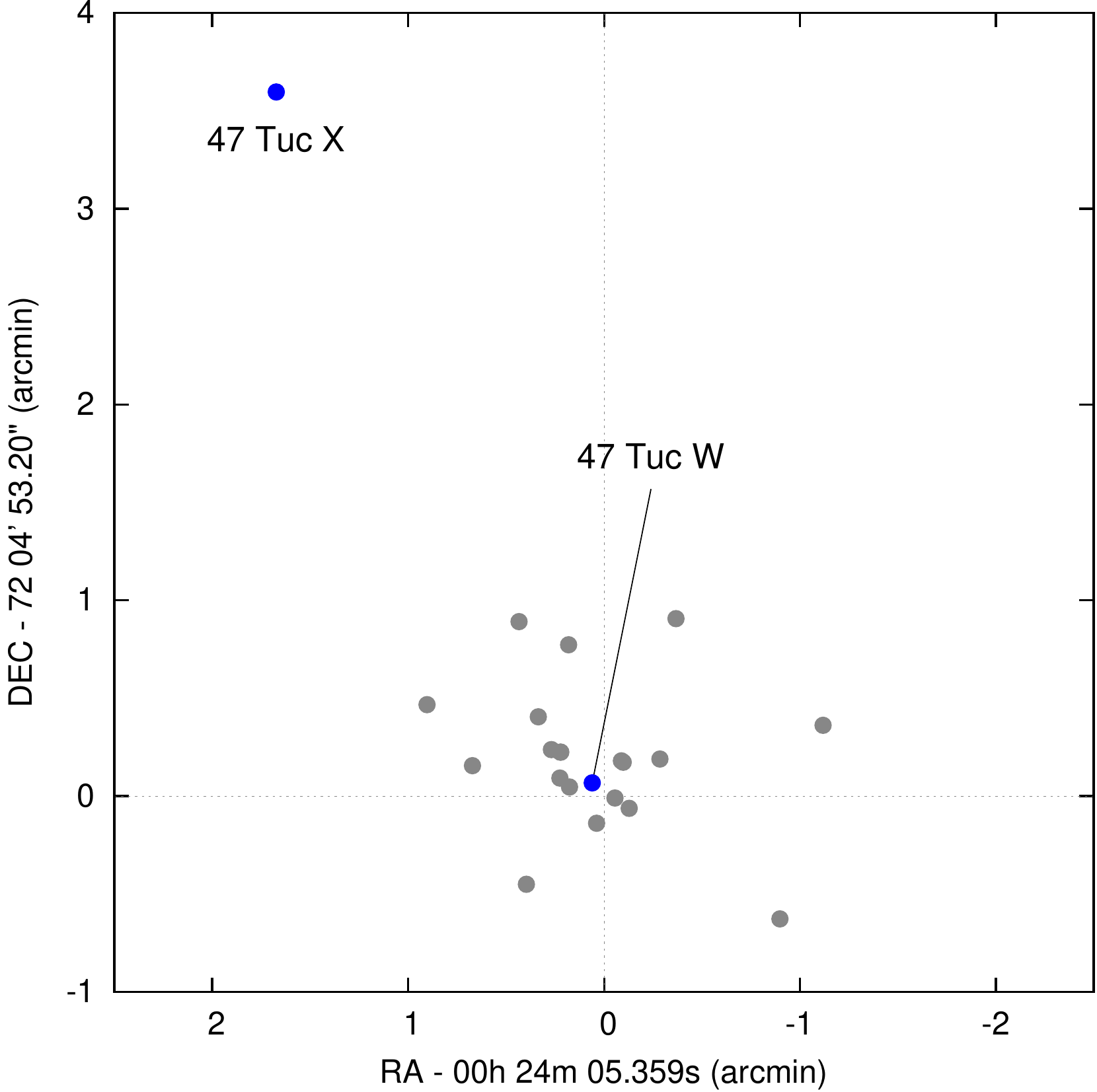}
   \caption{Position of 47 Tuc W and 47 Tuc X (blue dots) with respect to the other pulsars in the cluster (grey dots). The distance of \textcolor{black}{47 Tuc X} from the nominal cluster center is \mbox{$\sim$\,3.8 arcmin}, more than three times that of pulsar C, the second farthest in 47~Tuc.}
   \label{fig:pulsar_positions}
 \end{figure}
 
\citet{wsb+04} performed a photometric variability study of 47~Tuc in September 2002 (MJD 52508$-$52541), during the first gap in radio detections of 47~Tuc~V. 
They observed for 33 continuous nights, for 10~h/night with 6-min exposure times. The field of view was 52~arcmin per side, using a single filter that covered the Cousins $V$ and $R$ bands, and median seeing of 2.2~arcsec. They detected variables down to a magnitude of $V=20$, but blending prevented them from studying stars within the central 6 arcmin in radius. Two variables showed interesting periods of 0.2144 (V80) or 0.2155 (V34) days. V80, at 24 arcmin from 47 Tuc's center, should not be visible in the \textcolor{black}{14.4-arcmin wide beam of the Parkes radio telescope}, while V34, at 9.3 arcmin from 47 Tuc's center, is not too distant to be plausible, although the period is not an exact match. V34 is (barely) inside the field of view of \textcolor{black}{\emph{Chandra}} ACIS-I observations on March 16, 2000 (MJD 51619), and a \textcolor{black}{\emph{Chandra}} HRC-S observation on January 8, 2006 (MJD 53743), the latter also during the first gap in radio detections of 47~Tuc~V. Neither \textcolor{black}{\emph{Chandra}} observation showed a detection, with upper limits of $L_{\rm X}<3\times10^{30}$ \ergs\ (during 2000) and $L_{\rm X}<1\times10^{31}$ \ergs\ (during 2006). We rule out that any variable detected by \citet{wsb+04} is the counterpart of 47 Tuc V, and note that should 47 Tuc V have entered an active state in 2002, it would probably have been detected by this study.
\begin{figure*}
	{\begin{overpic}[width=\textwidth]{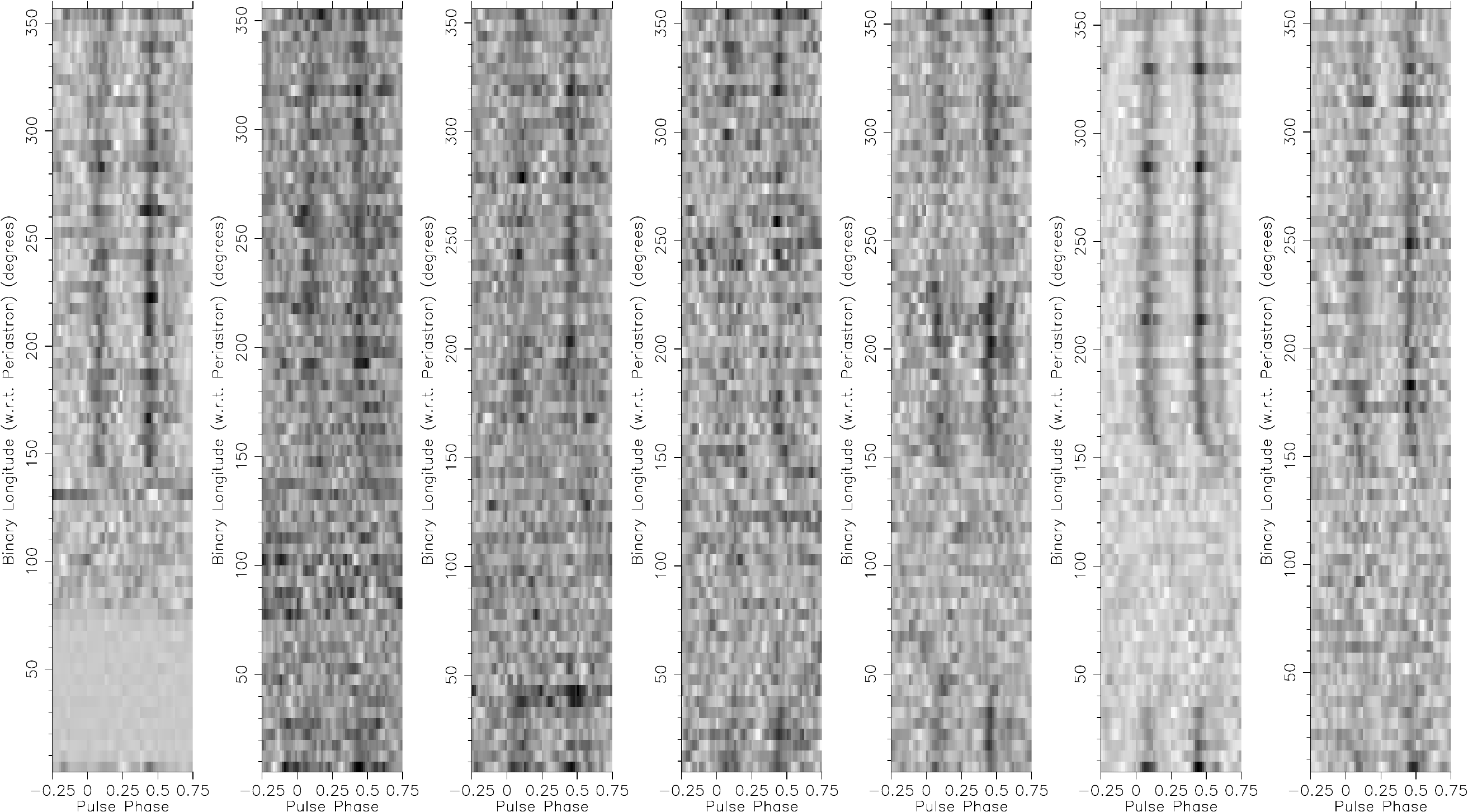}
\put (8.0,55.7) {\footnotesize{1}}
\put (22.3,55.7) {\footnotesize{2}}
\put (36.6,55.7) {\footnotesize{3}}
\put (50.8,55.7) {\footnotesize{4}}
\put (65.2,55.7) {\footnotesize{5}}
\put (79.5,55.7) {\footnotesize{6}}
\put (93.8,55.7) {\footnotesize{7}}
\end{overpic} }
   \caption{Intensity as a function of pulse phase ($x$-axis) and orbital phase $\phi_b$ ($y$-axis) for 7 sample detections of 47 Tuc W. The pulsar systematically exhibits eclipses between $\phi_b \simeq 0.09$ and $\phi_b \simeq 0.43$ that appear to be much more regular than those of 47 Tuc V.}
   \label{fig:eclipses_W}
 \end{figure*}
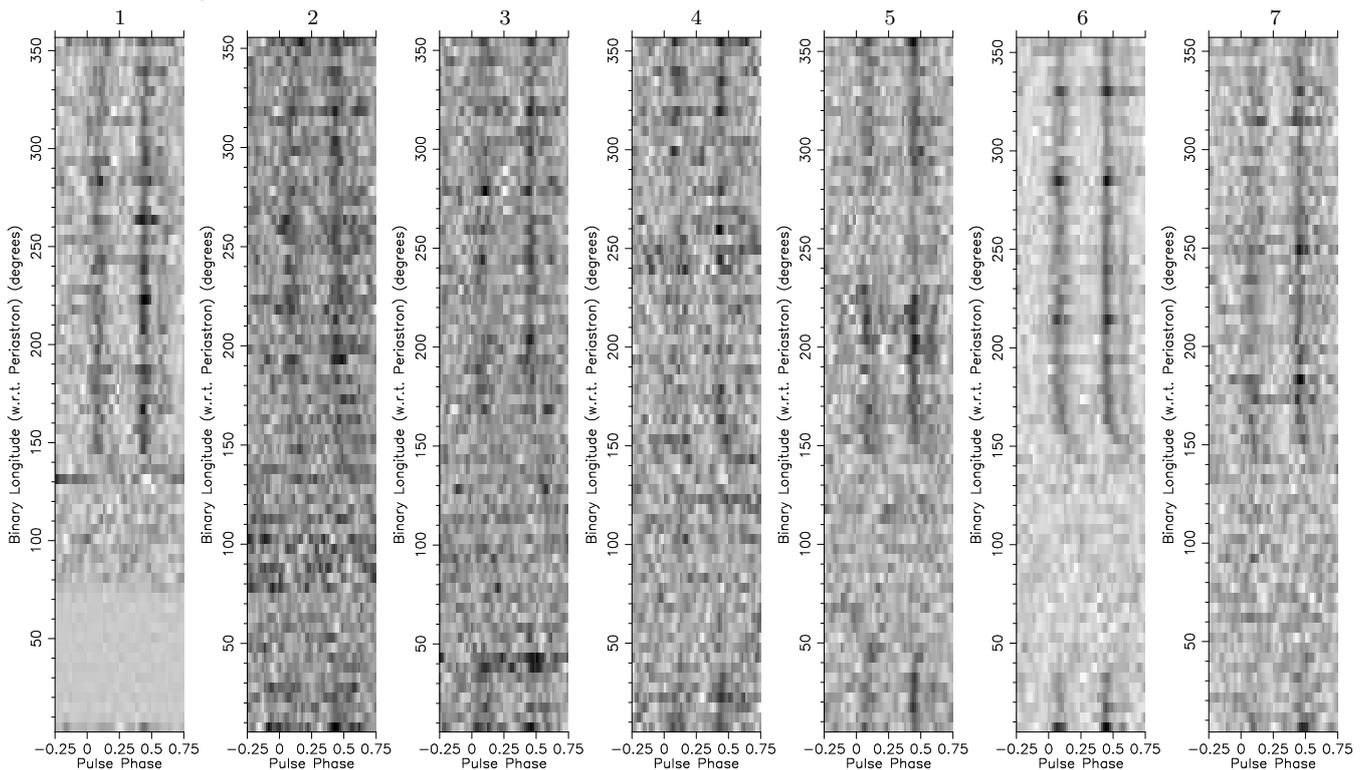
 
Finally, \citet{agb+01} performed a deep search for variable stars in a 8.3-day sequence of {\it HST} WFPC2 images, covering much of the central region of 47 Tuc (roughly 2.5 $\times$ 2.5 arcmin, off-centre). Due to the {\it HST} sharp point-spread function, this study was mostly complete down to $V=22$ for $>$\,10\% variations. No periodic signals were found that matched the period of 47 Tuc V. Since this study would have been sensitive to a signal from 47 Tuc V even in a passive state, given its predicted optical properties, it is likely that 47 Tuc V is not projected upon the central portion of the cluster investigated by \citet{agb+01}. We note that 47 Tuc W, with a shorter orbital period of 3.1 h, was identified in this dataset by \citet{egc+02} at $V=22.3$.

Observational evidence thus suggests that 47 Tuc V did not turn on as an accreting tMSP during its radio disappearances. Even bearing in mind the aforementioned caveats relative to the X-ray analysis, the lack of any compelling signals in the three optical surveys is a strong indication that 47 Tuc V is very likely not a tMSP.
\textcolor{black}{If so, the question about what is causing such long stretches of non-detections still holds. Scintillation can  be ruled out since it acts over much shorter timescales of hours or days in all 47 Tuc pulsars, and this was also the case for pulsar V during the intervals in which it was visible. A similar consideration applies to the observed eclipses, whose timescales are of the order of minutes for the short-lived events showed in Fig. \ref{fig:eclipses_V},  or hours for the regular obscurations seen during about half the orbit. We believe that some more fundamental and long-lived physical process, such as an increase in the rate of mass loss from the companion star, may have occurred. This may  have resulted in an engulfment of the pulsar, which thus became invisible until the mass loss switched back to the original rate.}

\subsection{47 Tuc W}
47 Tuc W is a 2.35-ms binary pulsar in a 3.2-h orbit. Like pulsars P and V, it was originally discovered by \citet{clf+00}, who detected it at a single epoch (MJD 51214) during which the favorable scintillation conditions brought the signal far above the detection threshold. On that occasion, the pulsar was also eclipsed for a large portion of the observation. The high S/N of the discovery observation allowed the authors to extract 12 ToAs and derive a first rough orbital solution, which was in turn used by \citet{egc+02} to obtain \textcolor{black}{the first optical identification of the companion of 47 Tuc W}. The association with the radio pulsar was validated by the optical photometry, which showed sinusoidal variations at a period consistent with that derived at radio frequencies. 

Our $T_0$-search produced 23 more detections of the radio MSP, from which we built a first incoherent timing solution. This was in turn used to refold the whole dataset, allowing us to spot the pulsar \textcolor{black}{in another 11 pointings}. Thanks to the large number of detections, we built a high-S/N template profile that revealed a previously unresolved third peak in the profile (Fig. \ref{fig:pulse_profiles}), which was instead blended with the major peak in \citet{clf+00}. With this template, we carefully extracted 187 more ToAs. Our timing data for 47 Tuc W thus consisted of 199 ToAs, spanning $\sim$\,10.2 years. The high number of ToAs, together with their frequent cadence, allowed us to obtain a phase-connected timing solution.
The timing residuals, shown in the top panel of Fig. \ref{fig:timing_residuals}, had a root mean square (r.m.s.) of 10.21 $\mu$s. The best-fit parameters are reported in Table \ref{tab:timing_solutions} and will now be discussed in detail.

\subsubsection{Astrometric parameters and dispersion measure}
Fig. \ref{fig:pulsar_positions} shows the radio timing position of 47 Tuc W relative to the nominal center of the cluster and to the other pulsars in 47 Tuc. 
With an angular distance of only 0.087 arcmin, it is the second closest pulsar to the cluster center. As expected, the measured right ascension ($\alpha$) and declination ($\delta$) are both consistent, within  $1.1\sigma$, with the optical position of the companion as measured by \citet{egc+02} on an {\em Hubble Space Telescope} astrometric frame tied to the positions of the other MSPs in the cluster. The measured proper motion along the same coordinates, $\mu_\alpha = 6.2 \pm 0.5$~mas~yr$^{-1}$ and $\mu_\delta = -2.5 \pm 0.3$~mas~yr$^{-1}$, is consistent with the global motion of the cluster, as calculated in Section \ref{sec:47TucP}, within less than $1\sigma$.
The DM was measured using the same method as for pulsars P and V and amounts to $24.367 \pm 0.003$ \dmunit, a value that, like in the case of 47 Tuc P, is very close to the average DM of all the 47 Tuc pulsars. According to the \citet{fkl+01} linear relation, this corresponds to a line-of-sight distance from the cluster center of just $d_\parallel \simeq -0.21$ pc. This corroborates the hypothesis, already suggested by the position, that the three-dimensional distance of 47 Tuc W from the cluster center is very small.

\subsubsection{Long-term orbital variability}
Also for 47 Tuc W, we needed to use a BTX model to correctly take into account the pulsar orbital motion. In this case the number of orbital frequency derivatives that we had to introduce was nine.
We also used the same method described in Section \ref{sec:orbital_variability_V} to study the long-term orbital variability. The resulting plot is shown in Fig. \ref{fig:orbital_variability}. The lower panel shows that the orbital period varies with an amplitude of a few milliseconds in a quasi-periodic fashion and with a characteristic timescale of roughly 3 years.
Contrary to the case of pulsar V, the coherence (phase-connection) of the timing solution and the frequent cadence of the data guarantee that the fitted BTX model is a reasonably faithful description of the actual changes in the orbital dynamics within the time span considered. However, the inability of the model to predict the orbital phase outside that range still holds. The large orbital variability prevents us from using 47 Tuc W as a probe for the cluster gravitational potential in the vicinity of the core.

\subsubsection{Eclipses and Redback nature}
The new 34 detections of 47 Tuc W constitute a good sample for a qualitative study of the eclipses. As shown in Fig. \ref{fig:eclipses_W}, the pulsar shows regular eclipses that last for about one third of its orbit. Using our sample of detections, we measured the mean eclipse ingress and egress orbital phases to be $\phi_b^\text{ing} = 0.09$ and $\phi_b^\text{egr} = 0.43$, respectively, resulting in the pulsar not being visible for $\sim$\,34\% of the orbit, on average. The eclipse is thus centered at phase $\phi_b = 0.26$, that is about 2 minutes later than the pulsar superior conjunction ($\phi_b = 0.25$). Also, the eclipse egress exhibits, on average, a smoother transition that is often accompanied by a delay of the pulses, whereas the ingress transition is generally shorter and more abrupt.
These characteristics are very similar to those found at X-ray wavelengths by \citet{bgv05}, who proposed that the X-ray eclipses could be caused by a shock produced by the interaction of the energetic pulsar wind with the gas spilling out from the companion via Roche-lobe overflow, through the inner Lagrangian point \citep[Fig. 2 in][]{bgv05}. The gas flow would have a ``cometary" shape due to the orbital Coriolis forces, from which the observed asymmetry of the X-ray and radio eclipses originate. Our radio observations are thus in support of this model.

\textcolor{black}{Based on the minimum companion mass of 0.127\,M$_\odot$, in combination with the presence of radio eclipses and of the strong orbital variability, 47~Tuc~W can be classified as a Redback system.}

\subsection{47 Tuc X}
To obtain the timing solution of 47 Tuc X, we followed an analogous procedure as for the case of pulsar W. After building a primitive ephemeris, we refolded our dataset to look for new detections and new ToAs to add to the \texttt{TEMPO} fit.
After converging to a phase-connected timing solution, we improved it by iterating the procedure a few more times, thus not only obtaining new detections (and hence, times of arrivals) but also improving the quality of the \textcolor{black}{previously extracted} ToAs.
The procedure led us to a total of 719 usable ToAs spanning $\sim$\,15 years.
The timing residuals, which are shown in the bottom panel of Fig \ref{fig:timing_residuals}, had a r.m.s. deviation of 14.51 $\mu$s. The best-fit parameters are reported in Table  \ref{tab:timing_solutions}.

\begin{figure}
  \includegraphics[width=\columnwidth]{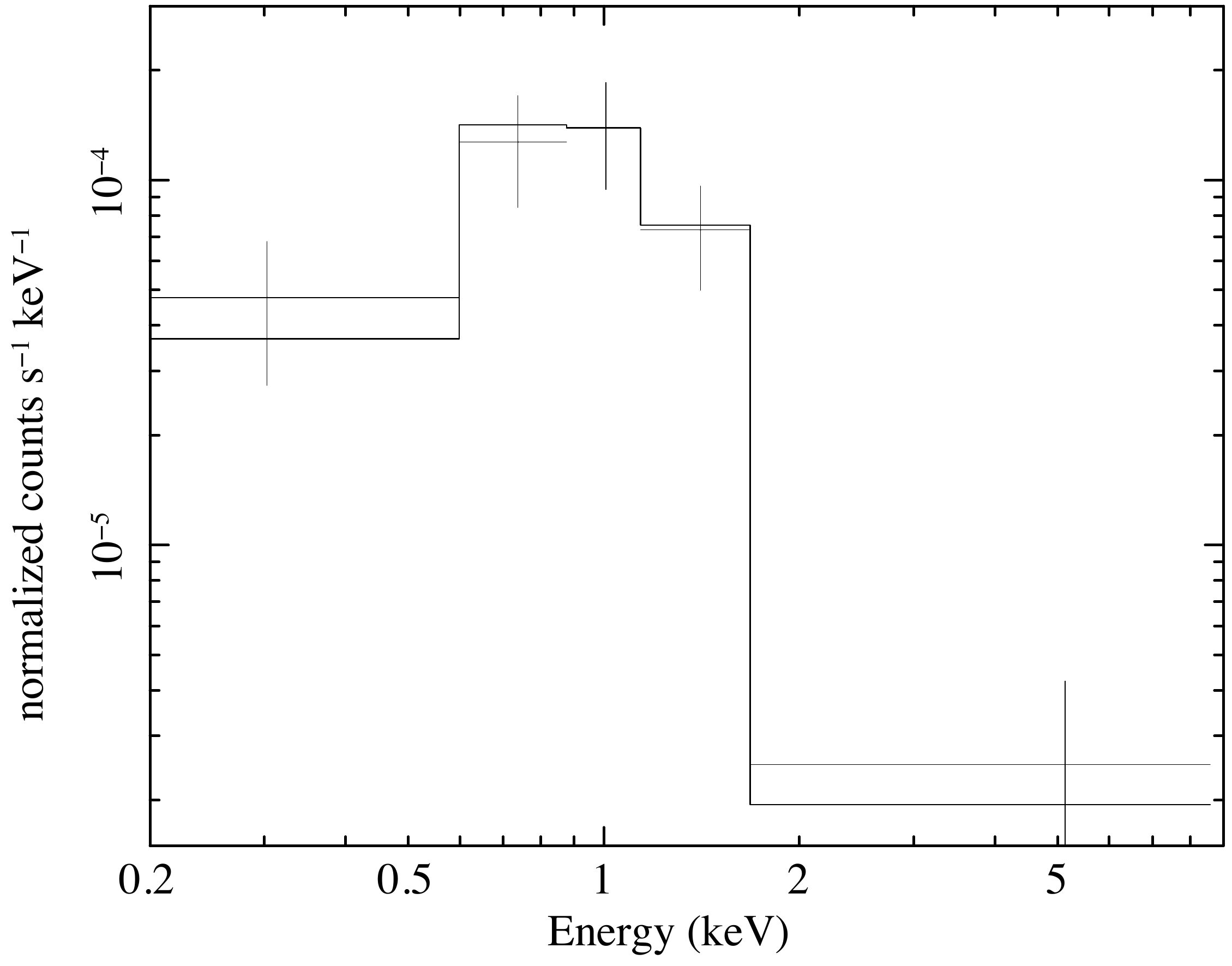}
  \caption{\textcolor{black}{\emph{Chandra}} X-ray spectrum of 47 Tuc X, fit with a hydrogen atmosphere neutron star model: observed data binned by 10 counts/bin (crosses), and best-fit model folded through instrumental response (solid line).}
  \label{fig:47TucX_chandra_spec}
\end{figure}

\subsubsection{Astrometric parameters and dispersion measure}
The first remarkable feature of 47 Tuc X is its position of $\simeq 3.83$ arcmin away from the cluster center, equivalent to $\simeq 10.63$ core radii and $\simeq 1.21$ half-mass radii\footnote{Assuming a core radius of 0.36 arcmin and a half-mass radius of 3.17 arcmin, as reported in the Harris GC catalogue (\url{http://www.physics.mcmaster.ca/~harris/mwgc.dat}).}. This is more than three times the distance of 47 Tuc C, the previous farthest pulsar known in 47 Tuc. Assuming a cluster distance of 4.69 kpc \citep{wgk+12}, this corresponds to a projected linear distance of $d_\perp \sim 5.22$ pc. 
Using the measured DM of $24.539 \pm 0.005$~\dmunit, and the linear relation by \citet{fkl+01}, we infer $d_\parallel \simeq 2.37$ pc. Such estimate has to be considered less robust than in the case of the other three pulsars, since the \citet{fkl+01} model is only known to work well in the central regions of the cluster.

The measured proper motion along the two coordinates are $\mu_\alpha = 5.8 \pm 0.1$~mas~yr$^{-1}$ and $\mu_\delta = -3.3 \pm 0.2$~mas~yr$^{-1}$. To these, we can subtract the global motion of the cluster as calculated in Section \ref{sec:47TucP}.
The motion of pulsar X relative to the cluster is therefore $\Delta \mu_\alpha = 0.9 \pm 0.9$~mas~yr$^{-1}$ in right ascension and of $\Delta \mu_\delta =  -0.6 \pm 0.7$~mas~yr$^{-1}$ in declination. Because of the large errors (which are dominated by the uncertainty on the GC proper motion) the range of possible trajectories of the pulsar in the cluster, obtained by numerically integrating its motion back in time, is too large to put any firm constraints on the dynamics of this system in 47 Tuc.

\subsubsection{Orbit and mass function}
Given the high circularity of the 47 Tuc X orbit and the absence of strong orbital variability, we opted for the ELL1 binary model \citep[][N. Wex, unpublished]{lcw+01}, which is particularly suitable for systems with very low eccentricities. 
The measured values for the first and second Laplace-Lagrange parameters are $\eta \equiv e \sin \omega =  (4.1 \pm 1.4) \times 10^{-7}$ and $\kappa \equiv e \cos \omega =  (-2.4\pm 1.5) \times 10^{-7}$, respectively, corresponding to an eccentricity $e = \sqrt{\eta^2 +\kappa^2 } = (4.8 \pm 1.5) \times 10^{-7}$.
The measured orbital period and projected semi-major axis of the pulsar orbit translate into a mass function $f(M_p) = 1.52 \times 10^{-2}$~\msun. For an assumed pulsar mass of 1.4 \msun, this implies a minimum ($i=90^{\circ}$) companion mass of $M_{\rm c}=0.36$~\msun\ and a median ($i=60^{\circ}$) value of $M_{\rm c}=0.43$~\msun.

 \subsubsection{X-ray detection}
 \label{sec:x_ray_47TucX}
47 Tuc has been studied repeatedly by the \textcolor{black}{\emph{Chandra X-ray Observatory}}, but no list of \textcolor{black}{\emph{Chandra}} X-ray sources outside the half-mass radius has been
published.  We utilize the deepest \textcolor{black}{\emph{Chandra}} observations of 47 Tuc that cover the position of 47 Tuc X and retain spectral information.  These are 4
$\sim$\,65-kilosecond observations done in 2002 using the ACIS-S array in full-frame mode, described by \citet{hge+05} (the short
interleaved subarray observations do not cover the position of 47 Tuc X). Note that these observations, with a \textcolor{black}{3.2-s} frame readout time, are not
sensitive to pulsations at the X-ray spin period.
We downloaded the Level 2 reprocessed event lists provided by the \textcolor{black}{\emph{Chandra}} X-ray Center pipeline reprocessing\footnote{\url{http://cxc.harvard.edu/cda/repro4.html}}, which applied the latest calibration files to the data. We reprojected the 4 event files to a common tangent point, constructed an image of the S3 CCD chip in the 0.3$-$6 keV energy band, and ran the \texttt{wavdetect} source detection algorithm.  

An X-ray source is clearly detected at the coordinates $\alpha=00$:24:22.416, $\delta = -$72:01:17.29, with astrometric uncertainty of 0.6 arcsec (90\% confidence), consistent with the radio timing position of 47 Tuc X.  At this distance from the cluster core (3.83 arcmin) source crowding is low, with a 0.1\% probability of a chance coincidence based on the local density of detected sources, so we are confident that this is the true X-ray counterpart. 

We extracted source and local background spectra and constructed response files for each of the 4 observations, and combined them.  Only 35 counts are attributed to 47 Tuc X, \textcolor{black}{after subtracting a local background of 12.3$\pm$4.6 counts (the low count rate and relatively high background are due in part to 47 Tuc X's position significantly off-axis, reducing the detector sensitivity and increasing the size of the point spread function)}. Therefore, we grouped the spectra by 10 counts/bin (retaining the last underfilled bin) and \textcolor{black}{fitted} only simple spectral models with the \texttt{XSPEC} X-ray spectral fitting program \citep{a96}. In each fit, we included photoelectric absorption by the interstellar medium (the \texttt{XSPEC} model \texttt{tbabs}) using \citet{wac00} abundances and \citet{vfk+96} cross-sections, with the hydrogen column density, $N_{\rm H}$, fixed to $1.3\times10^{20}$~cm$^{-2}$  \citep{gbc+03,ps95}.  We also performed fits to the unbinned spectra using the C-statistic \citep{c79}, with consistent results.

A simple absorbed power-law fit is statistically acceptable, with an inferred photon index of 2.4$^{+0.5}_{-0.4}$. Such a soft value for the photon
index would be unusual for magnetospheric emission from radio pulsars, where the index is typically between 1 and 2 \citep[e.g.][]{ba02}.  It is also rather softer than the photon index (1.1$-$1.7) measured in cases where the emission is thought to be powered by a shock between a wind from the donor star, and a pulsar wind \citep[e.g.][]{bgv05,bvh+10}. Thus, it seems likely that part or all of the X-rays are produced by emission from one or both heated polar caps, as seen in many MSPs with similar spin properties and X-ray luminosities \citep{zps+02,z06,bgh+06}.   We therefore \textcolor{black}{fitted} the spectrum with blackbody (for consistency with previous works) and hydrogen neutron star atmosphere (NSATMOS, \citealt{hrn+06}) models, both of which were good fits. The X-ray spectrum with the fitted NSATMOS model is shown in Fig. \ref{fig:47TucX_chandra_spec}. For the blackbody fit, the inferred temperature is $3.0^{+1.0}_{-0.8}\times10^6$ K with an inferred equivalent radius of 0.07$\pm0.05$ km. For the NSATMOS model, we fixed the neutron star mass to 1.4 \msun, the radius to 10 km, and the distance to 4.69 kpc \citep{wgk+12,hka+13}, and allowed the normalization to vary.  In this case, the inferred temperature is $1.9^{+1.0}_{-0.7}\times10^{6}$ K, and the inferred radius is 0.2$^{+0.4}_{-0.1}$ km.  In either case, the inferred 0.5$-$10~keV luminosity is $2.0\pm0.6\times10^{30}$ erg~s$^{-1}$.

Comparing the spectral fits to the X-ray spectrum of 47 Tuc X with those for other MSPs, we see that the best-fit temperatures tend to be higher than those for the other
MSPs in 47 Tuc \citep{bgh+06}, while the best-fit emitting radii are smaller.  However, 3 of the 5 MSPs in NGC 6752 have higher fitted blackbody temperatures, and smaller best-fit radii, than 47 Tuc X \citep{fhc+14}.  The remarkably high temperature of 47 Tuc X, as for those in NGC 6752, could be due to the presence of an undetected non-thermal spectral component, in addition to the predominating thermal component. The limited statistics preclude a definite answer.

\begin{figure}
\includegraphics[width=0.9\columnwidth]{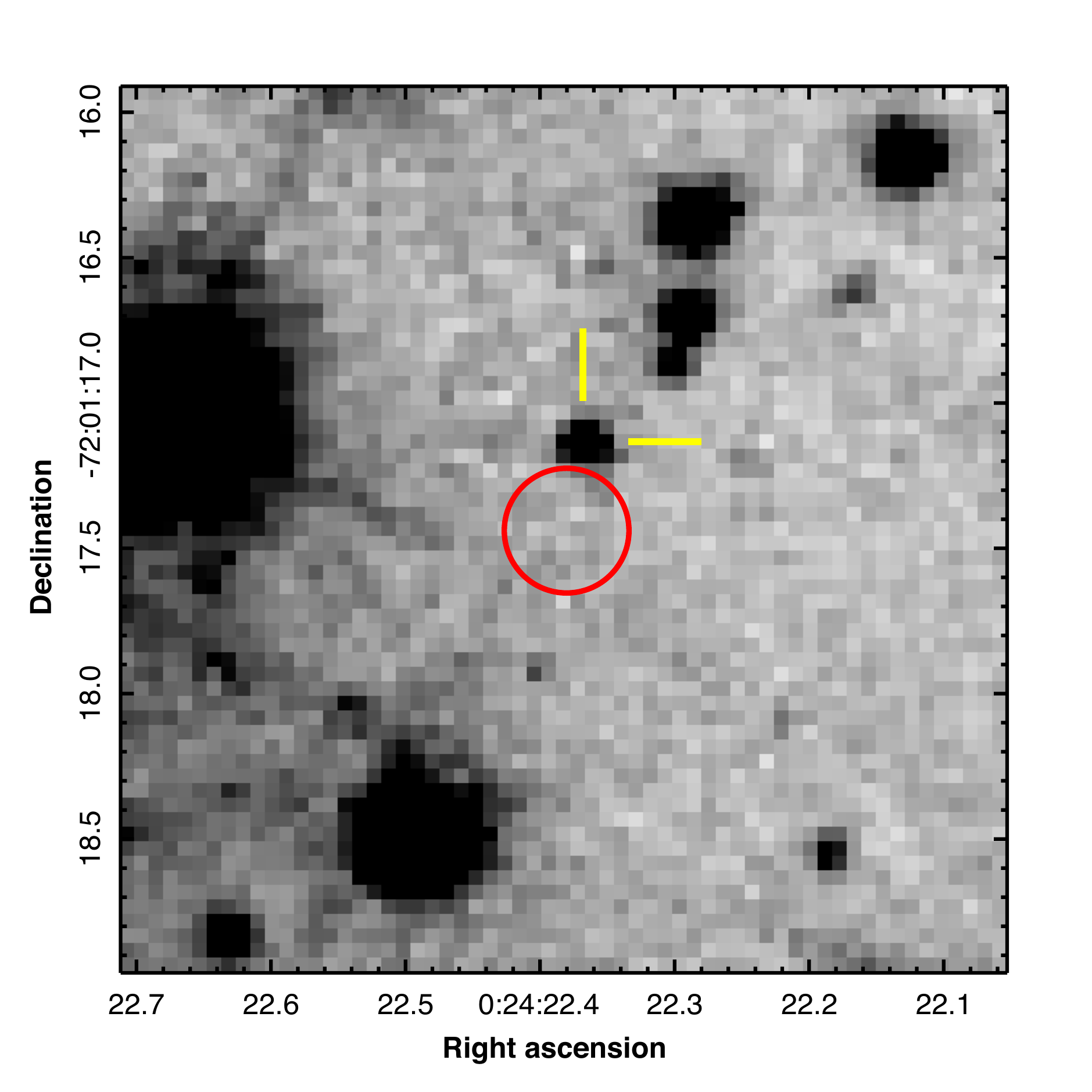}
\caption{A section of the stacked ACS/WFC F435W image centered on the radio position of 47 Tuc X, which is marked with a red circle. The radius of the circle is equal to the 3$\sigma$ positional error (0\farcs21). The star that is indicated with tick marks is the closest detection to the radio position at a separation of 0\farcs3. The size of the image is 3\arcsec $\times$ 3\arcsec. North is up, east to the left. \label{fig_fc}}
\end{figure}

\subsubsection{Characteristic age, magnetic field and spin-down luminosity}
Our timing precision was enough to detect the orbital period derivative, which amounts to $\dot{P_b} = 6 \pm 2 \times 10^{-12}$ s\,s$^{-1}$. This opens the possibility of constraining the intrinsic spin down of the pulsar, and thus other relevant quantities that depend upon it.
As discussed in \citet{p93} and \citet{lk05}, a variation of the orbital period can be the result of multiple effects, many of which contribute equally also to the observed spin period derivative, $\dot{P}$. Others, like the gravitational wave damping, are much smaller and can be neglected. In the end, for a system like 47 Tuc X, we can relate the intrinsic pulsar spin-down, $\dot{P}_\textrm{int}$ to only observed quantities:

\begin{equation}
\dot{P}_\textrm{int} = P_\textrm{obs} \Bigg [ \bigg ( \frac{\dot{P}}{P} \bigg )_\textrm{obs} - \bigg ( \frac{\dot{P_b}}{P_b} \bigg )_\textrm{obs}\Bigg ]
\end{equation}
where we approximated the intrinsic spin period, $P_\textrm{int}$, with the observed one, $P_\textrm{obs}$. We obtain $\dot{P}_\textrm{int} = -1 \pm 3 \times 10^{-20}$ at the $3\sigma$ level. Because the pulsar must be spinning down, the real value must be positive. Hence we can put an upper limit and say that $\dot{P}_\textrm{int} \lesssim 2 \times 10^{-20}$ with $\sim$\,99\% probability.

From this we can also give limits for the spin-down luminosity of the pulsar, $L_\textrm{sd}$, its characteristic age, $\tau_c$, and its surface magnetic field, $B_s$ using equations (3.5), (3.12) and (3.15), respectively, from \citet{lk05}.
We find $L_\textrm{sd} \lesssim 7 \times 10^{33}$ erg~s$^{-1}$, $ B_s \lesssim 3 \times 10^{8}$ G and $\tau_c \gtrsim 4$ Gyr at the $3\sigma$ level.

\begin{figure}
\includegraphics[width=0.9\columnwidth]{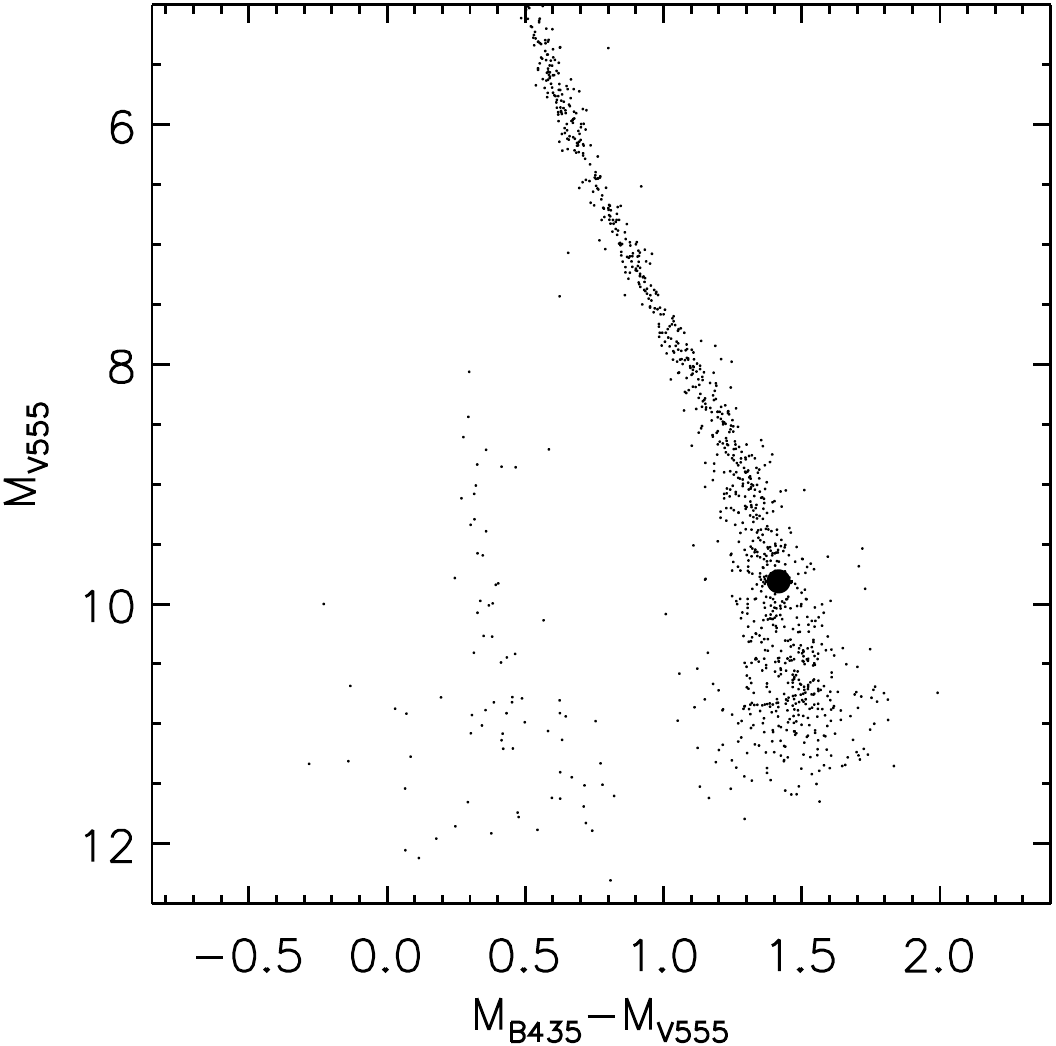}
\caption{$B_{\rm 435} - V_{\rm 555}$ versus $V_{\rm 555}$
  color-magnitude diagram extracted from a $\sim$\,1 arcmin$^2$ section of the ACS/WFC images around 47 Tuc X. The nearest detection to 47 Tuc X is marked with a large filled circle, and has typical main-sequence$-$star colors. Photometry is given in the Vega-mag system and has been converted to absolute magnitudes using an unreddened distance modulus of $(m-M)_0 = 13.36$ \citep{wgk+12} and a reddening of $E(B-V)=0.024$ \citep{gbc+03}. Errors on the
  photometry are comparable to the symbol size, and include the {\sc DAOPHOT} errors, and the errors on $(m-M)_0$ and $E(B-V)$. The sequence of stars with $M_{V_{\rm 555}} \gtrsim 8$ and $0.2 \lesssim M_{B_{\rm 435}}-M_{V_{\rm 555}} \lesssim 0.7$ are stars in the Small Magellanic Cloud. \label{fig_cmd}}
\end{figure}

\subsubsection{Search for the optical counterpart}
\label{sec:optical_search}

The location of 47 Tuc X, being almost 4~arcmin away from the center of 47 Tuc, was observed with  {\em HST}
on only one occasion. The four images that include the position of 47 Tuc X were taken with the Wide Field Channel (WFC) on the Advanced Camera for Surveys (ACS) as part of program GO-12971 (PI: H. Richer). These observations were obtained on 2013 February 26 between 05:32 and 7:19 UT, and include two exposures (one of $290$~s and one of $690$~s) in the F435W filter, and another two (of $360$~s and $660$~s) in the F555W filter. We retrieved these images from the {\em HST} archive and started our data reduction with the flat-fielded {\tt flc} images, which are corrected for the degradation of the charge transfer efficiency by the CALACS pipeline version 8.3.0. We processed the images with the \texttt{DRIZZLEPAC} software \citep{ghf+04} to remove the geometric distortion and create stacks for each filter. The resulting master images have a pixel scale of $0.05$~\mbox{arcsec pixel$^{-1}$}. Next, we tied the absolute astrometry to the same reference system as the radio coordinates, i.e. the International Celestial Reference System (ICRS), using stars in the UCAC2 catalog \citep{zuz+04}. As the UCAC2 stars in the ACS images are all saturated, we employed the following method. Based on the cataloged coordinates of 140 UCAC2 stars, we first derived an astrometric solution for a 30-s $V$-band image of the cluster taken on 2002 October 29 with the Wide Field Imager on the ESO/2.2-m telescope at La Silla, Chile. The resulting solution has r.m.s. residuals of $\sim$\,~0\farcs026 in both right ascension and declination. From this image we selected 24 unsaturated and relatively isolated stars in the vicinity of 47 Tuc X to astrometrically calibrate the 290-s F435W exposure of GO-12971, resulting in a r.m.s. residuals of 0\farcs035 in right ascension and 0\farcs045 in declination. Finally, we transferred this solution to the stacked F435W and F555W images of the field around 47 Tuc X with $\sim$\,500 stars and negligible r.m.s. residuals ($\sim$\,0\farcs01). We used \texttt{DAOPHOT} to extract photometry; the resulting magnitudes in the F435W and F555W filters (denoted with $B_{\rm 435}$ and $V_{\rm 555}$, respectively) were calibrated to the Vega-mag system using the zeropoints provided by the STScI\footnote{\url{http://www.stsci.edu/hst/acs/analysis/zeropoints}}.
Fig.~\ref{fig_fc} shows a section of the F435W image around the radio position of 47 Tuc X, which is marked with a circle. The radius of the circle represents the 3$\sigma$ error on the absolute astrometry of the image. Here, $\sigma=0\farcs072$ and equals the quadratic sum of all the astrometric errors mentioned above, plus the systematic error in the alignment of the UCAC2 coordinates to the ICRS
(about 0\farcs010; \citealt{zuz+04}). The errors on the radio position are negligible. The star that is closest to 47 Tuc X is located at a separation of 0\farcs3 (4.1$\sigma$).  In the color-magnitude diagram of Fig.~\ref{fig_cmd}, this star falls right on the main sequence. We consider its angular offset from 47 Tuc X too large to make this star a convincing counterpart. Also, the evolutionary path for such a companion would exclude such a low eccentricity as we measure (see below). For example, the He WD companions of five 47 Tuc MSPs that \cite{rvh+15} identified with the same method as used here, were all excellently aligned to the radio positions, with a maximum offset of
0.016~arcsec (0.2$\sigma$). 
From the faintest detection with ${\rm S/N} \gtrsim 3$ we
estimated the detection limit in the F435W band. We find that it corresponds to an absolute magnitude of $M_{B_{\rm 435}}\simeq13.7$, where we adopted the unreddened distance
modulus $(m-M)_0 = 13.36\, \pm\, 0.02\, \pm\, 0.06$ (random and systematic error, respectively) from \cite{wgk+12} and the reddening $E(B-V)=0.024\pm\,0.004$ from \cite{gbc+03}. In the standard MSP formation scenario, the companion to the MSP is a low-mass WD, which is all that is left of the original donor star after it has lost most of its envelope during the mass-transfer stage. Based on the theoretical relation between the WD mass and the orbital period of the MSP$-$WD binary, the measured orbital period of 10.92 days for 47 Tuc X implies a WD mass between $\sim$\,0.22 and 0.27 \msun \citep{ts99,db10}. Based on our non-detection of a WD near 47 Tuc X and the cooling tracks of 0.2$-$0.3 \msun WDs computed for the metallicity of 47 Tuc \citep[see][]{rvh+15}, we find that the lower limit to the age of the WD must be $\sim$\,1.7\,Gyr for it to remain undetected, assuming the companion is indeed a low-mass WD.

\begin{table*}
\centering
\begin{tabular}{cclccc}
\hline
Group 			& \# of pulsars		&Pulsar names 			& $P_b$ (days) 		& $M_{\rm c}$ (\msun)\\
\hline
Isolated			&10			&C, D, F, G, L, M, N, Z, aa, ab  	&-			&-			\\
Short-$P_b$		&8			&I, J, O, P, R, V, W, Y	&0.06$-$0.4		&$\sim$\,0.03			\\
Long-$P_b$		&6			&E, H, Q, S, T, U		&0.4$-$2.36		&$\sim$\,0.2			\\
Very-Long-$P_b$		&1			&X				&10.92				&$>$\,0.36			\\
\hline
\end{tabular}
\caption{Population of pulsars in 47 Tuc.}
\label{tab:pulsar_groups}
\end{table*}

\subsubsection{Formation}
Ranking 3rd among the most circular systems with a measurable non-zero eccentricity (after PSR J1909$-$3744 and PSR J1738+0333), the circularity of 47 Tuc X is even more remarkable when related to its long orbital period and to the fact that it resides in a globular cluster. Indeed, pulsar X is by far the most circular system ever found in a GC and, more generally, it is the binary with the lowest eccentricity-to-orbital-period ratio known, with a value of $e/P_b = 4.4 \times 10^{-8}$~days$^{-1}$ (compared to $7.4 \times 10^{-8}$~days$^{-1}$ of PSR J1909$-$3744 and $9.6 \times 10^{-7}$~days$^{-1}$ of PSR J1738+0333).
The eccentricity and the position of 47 Tuc X can give us important clues about how this system may have formed. Here we discuss two possible formation hypotheses, where the main difference is the orbit of the system in the cluster.

In the first scenario, an unrecycled NS near the core of the GC has a close encounter with a main sequence (MS) binary, with subsequent chaotic interactions. In such encounters, the most likely outcome is the high-speed ejection of the lightest star among the three (likely one of the light MS stars); by conservation of
momentum, the newly formed NS$-$MS binary will also recoil. Its orbit around the cluster will become eccentric, and it will spend most of its time in the outskirts of the cluster. Such an exchange interaction would induce a residual eccentricity. The extremely low eccentricity of the orbit implies that the circularization occurred \emph{after} the formation of the current binary, hence the current companion star must be the former donor that spun up the pulsar. This is similar to the case of pulsar PSR J1911$-$5958A in NGC 6752 \citep{bvk+06,cbp+12}.
We also note that, if the system has actually been ejected from the core but is still bound to the cluster, it will be making many more periodic visits to the central regions of 47 Tuc. Even though the relaxation \textcolor{black}{timescale} at the projected distance of 47 Tuc X is more than 3~Gyr, because the orbits within the cluster are non-Keplerian, the system will spend a fairly long  time in proximity of the core, thus significantly reducing the relaxation \textcolor{black}{timescale} and, equivalently, increasing the probability of dynamically interacting with other bodies. This would in turn increase the eccentricity of the binary \citep{p92}. Because 47 Tuc X instead has an extremely circular orbit, the system must be relatively young.
There is nothing unusual in this formation path $-$ this is the normal evolution for all MSPs in globular clusters. Most other systems formed in this way eventually sink back to the core of the cluster, because of dynamical friction, and eventually reach an equilibrium configuration dictated by mass segregation \citep{hge+05}. This is the reason why all other MSPs apart from 47 Tuc X lie, at least in projection, within 1.2 arcmin from the center of the cluster. The fact that 47 Tuc X is found in the outskirts would suggest that, since recycling, it has had no time to sink back to the core, which would in turn indicate that it was recycled later than the other MSPs in the cluster.

However, despite these two indications that the system must be young, the optical non-detection introduced a lower limit on the age of the system of $\sim$\,1.7 Gyr.
This problem is avoided in the second possible formation scenario, where the system was born directly in the outskirts. In this case, 47 Tuc X may have formed in two ways: either from a primordial MS$-$MS star that naturally evolved first into a LMXB and then into the current pulsar$-$WD binary we see today, or again from a dynamical encounter that set the system to a nearly circular orbit around the cluster. In both ways, the motion in the cluster would be such that the system never approached the central regions, the probability of dynamical encounters was low and thus the evolution of 47 Tuc X would resemble that which happens in the Galactic field, which naturally \textcolor{black}{retains} the low eccentricity \textcolor{black}{obtained} after the end of the recycling phase. 

In both scenarios, it is difficult to tell whether the companion is a He WD or a Carbon-Oxygen (CO) WD. The median mass of the companion, inferred from the mass function, $M_{\rm WD}\simeq 0.43$~\msun, lies roughly at the border that discriminates the two types. However, as already mentioned at the end of Section \ref{sec:optical_search}, the companion mass range predicted by the $P_b-M_{\rm WD}$ relation for He WDs \citep{ts99} is somewhat lower than the minimum mass derived from the mass function, thus making a CO WD type more likely.

Regardless of the formation history and the nature of its companion, 47 Tuc X does not belong to any of the groups of pulsars so far discovered in the cluster (see Table \ref{tab:pulsar_groups}). However, this is quite understandable: its position in the outskirts of 47 Tuc enables it to maintain a long orbital period for a long time without binary destruction, while the other pulsars, closer in, would likely be ionized (thus leaving isolated MSPs) by stellar encounters. 47 Tuc X looks just like many other MSPs in the field, which are in wide orbits together with a He WD companion. The dense and ionizing dynamical environment, generally prevents us from seeing such systems in globular clusters.

\section{Conclusions}
\label{sec:conclusions}
We have presented a study of four binary millisecond pulsars belonging to the globular cluster 47 Tucanae that, despite being known, had never been investigated in detail because of the sparsity of their detections.
We have used different search techniques on 16 years of archival data, which allowed us to detect each pulsar at many more epochs. Thanks to these, we were able to time and thus better characterize the pulsars.
For two pulsars, namely 47 Tuc P and V, the detection rate was still too low to allow us to obtain a phase-connected timing solution, but we were able to obtain very precise measurements of their spin periods, DM's and orbital parameters. 47 Tuc V is a Redback-like  binary system, eclipsed for 50\% of its orbit. Although its characteristics and behaviour are quite similar to those of the so-called ``transitional" MSPs, we showed that it seems unlikely that   47 Tuc V actually belongs to the latter class.
For 47 Tuc W and X, we successfully obtained phase-connected timing solutions that allowed a full characterization of both systems. The former is a well-known Redback that has already been extensively studied at other wavelengths. Our radio timing complemented those studies and revealed a strong orbital variability, as it is often the case for Redbacks.
47 Tuc X is probably the most peculiar system and stands out for its long orbital period, its extreme circularity, as well as its distance from the cluster center, compared to all the other pulsars in 47 Tuc. We have studied this system in the radio, X-ray and optical wavelengths and discussed its possible formation.

\textcolor{black}{The results presented in this work allowed a more complete characterization of the pulsar population of 47 Tuc. The properties of this population largely conform to the expectations of \citet{vf14} for a cluster with a low rate of stellar interactions per binary. In such a cluster, exchange encounters form NS$-$low-mass MS star binaries; these are the only type of MS star remaining in any globular cluster. The slow evolution of such stars and resulting long (and, in the case of 47 Tuc, undisturbed) accretion episode then lead invariably to fast-spinning pulsars (as observed in this and other lower density clusters) in systems similar to the binary MSPs in the Galaxy: MSP$-$He WD systems, Redbacks, Black-Widows and (perhaps evolving from the latter) isolated MSPs. Indeed, none of the binaries found so far in 47 Tuc is clearly the result of an exchange interaction. The only possible exception to this is 47~Tuc~X, an unusual system that could either have resulted from a primordial binary, or could have formed in an exchange encounter from a population of neutron stars that had not yet been segregated into the core of the cluster.}

In the last few years, new and more sensitive observations of 47 Tuc have been carried out with the Parkes radio telescope. Even though it is unlikely that these new data will allow us to obtain a phase-connected solution for 47 Tuc P in the short term, the chances that this could be the case for pulsar V look higher. If so, this could help confirm or reject the hypothesis of a non-transitional nature for the object. For 47 Tuc X, the extended data span could improve the measurement of the orbital decay and the proper motion.
A possible major leap in the study of these four pulsars might soon be possible with the MeerKAT\footnote{\url{http://public.ska.ac.za/meerkat}} telescope \citep[][]{bj12}, which,  thanks to its position in the southern hemisphere, will allow a study of 47 Tuc at radio wavelengths with an unprecedented sensitivity.

\section*{Acknowledgements}
A. R. and P. F. gratefully acknowledge financial support by the European Research Council for the ERC Starting grant BEACON under contract No.
279702. A. R. and P. T. are members of the International Max Planck research school for Astronomy and Astrophysics at the Universities of Bonn and Cologne. A. R. acknowledges partial support through the Bonn-Cologne Graduate School of Physics and Astronomy.
C. H. acknowledges support from an NSERC Discovery Grant and an Alexander von Humboldt Fellowship, and is grateful for the hospitality of MPIfR. C. G. B. acknowledges support
from the European Research Council under the European Union's Seventh Framework Programme (FP/2007-2013) / ERC Grant Agreement nr. 337062 (DRAGNET; PI: Jason Hessels).
We especially thank Ralph Eatough for observations, significant assistance with the data handling and for stimulating discussions on pulsar searches. A. R. also thanks Golam Shaifullah for nice and useful discussions.




\bibliographystyle{mnras.bst}
\bibliography{Ridolfi_Paper_47_Tuc_Binaries_revision1.bib} 

%
%
%
%
%


\bsp	
\label{lastpage}
\end{document}